\documentclass[sigconf,nonacm]{acmart}

\usepackage{acmart-taps} % TAPS-specifics
\usepackage{subcaption}
\usepackage{hyperref} % set links
\usepackage{caption} % set captions
\usepackage{dirtytalk} % quotes
\usepackage{seqsplit}

\usepackage{multicol} % Multi columns
\usepackage[dvipsnames]{xcolor} % custom colors and named color package

\usepackage{array} % flexible column formatting, fixes spacing
\usepackage{booktabs} % better vertial lines/rules

\usepackage{todonotes}

\usepackage{tikz}
\usetikzlibrary{positioning,arrows.meta,calc}
\definecolor{deepteal}{HTML}{0B6A72}
\definecolor{wordteal}{HTML}{187868}
\definecolor{coral}{HTML}{E0765C}
\definecolor{ink}{HTML}{35484A}
\definecolor{muted}{HTML}{7C8E8E}
\definecolor{hair}{HTML}{D5E4E2}
\definecolor{cardbg}{HTML}{FBFEFD}

\usepackage{xcolor}
\definecolor{trendup}{RGB}{0,158,115}    % bluish green
\definecolor{trenddown}{RGB}{213,94,0}   % vermillion
\definecolor{trendflat}{RGB}{110,110,110}% gray
\newcommand{\tup}{\textcolor{trendup}{$\uparrow$}}
\newcommand{\tdown}{\textcolor{trenddown}{$\downarrow$}}
\newcommand{\tflat}{\textcolor{trendflat}{$\rightarrow$}}
\newcommand{\tupdown}{\textcolor{trendup}{$\uparrow$}\!\textcolor{trenddown}{$\downarrow$}}
\newcommand{\tdownup}{\textcolor{trenddown}{$\downarrow$}\!\textcolor{trendup}{$\uparrow$}}

\AtBeginDocument{%
  }

\setcopyright{acmlicensed}
\copyrightyear{2027}
\acmYear{2027}
\acmDOI{}
\acmConference[CHI '27]{CHI Conference on Human Factors in Computing Systems}{May 10-14, 2027}{Pittsburgh, PA}
\acmISBN{}

\begin{document}

% \title[Writing at CHI]{Writing at CHI: How the Prose Changed Throughout the History of the CHI Conference}
% \title[How Did Writing Change At CHI?]{How Did Writing Change At CHI? Comparing CHI Writing Before And After The Large Language Model Period}
\title[How Did Writing Change At CHI?]{How Did Writing Change At CHI? Analyzing 44 Years of CHI Writing Before and After the Introduction of Large Language Models}

\author{Thomas Kosch}
\email{thomas.kosch@hu-berlin.de}
\orcid{0000-0001-6300-9035}
\affiliation{%
  \institution{HU Berlin}
  \city{Berlin}
  \country{Germany}
}

\author{Robin Welsch}
\email{robin.welsch@aalto.fi}
\orcid{0000-0002-7255-7890}
\affiliation{%
  \institution{Aalto University}
  \city{Espoo}
  \country{Finland}
}

\author{Michael Hedderich}
\email{hedderich@cis.lmu.de}
\orcid{0000-0001-6858-0791}
\affiliation{%
  \institution{LMU Munich, MCML}
  \city{Munich}
  \country{Germany}
}

\author{Christopher Katins}
\email{christopher.katins@hu-berlin.de}
\orcid{0000-0001-6257-7057}
\affiliation{%
  \institution{HU Berlin}
  \city{Berlin}
  \country{Germany}
}

% What is the problem?
% Why is the problem important?
% What is the solution?
% What is the approach?
% What are the findings?
% What are the implications in the bigger picture?

\begin{abstract}

The availability of Large Language Models (LLMs) reshaped scientific discourse at a linguistic level. LLMs are assumed to homogenize academic writing, flattening it into a single generic lexical register. To understand how CHI writing has changed since the public release of LLMs, we analyzed full texts of 14,262 archival papers across all 44 CHI proceedings from 1982 to 2026, measuring readability, register, lexical diversity, and marker words typically produced by LLMs. We find that prose did not homogenize, while vocabulary grew more varied, and sentence rhythm remained irregular. CHI prose changed more between 2016 and 2026 than in other decades toward greater density, and reading ease has declined since 2022. The word-level shift began before any author used LLMs, so LLMs did not start the change but accelerated it. Reflecting on the history of CHI papers, we discuss what may have caused changes in prose and how LLMs accelerated them.

\end{abstract}

\begin{CCSXML}
<ccs2012>
   <concept>
       <concept_id>10003120.10003121</concept_id>
       <concept_desc>Human-centered computing~Human computer interaction (HCI)</concept_desc>
       <concept_significance>500</concept_significance>
       </concept>
 </ccs2012>
\end{CCSXML}

\ccsdesc[500]{Human-centered computing~Human computer interaction (HCI)}
\keywords{Lexical Analysis, Large Language Models}

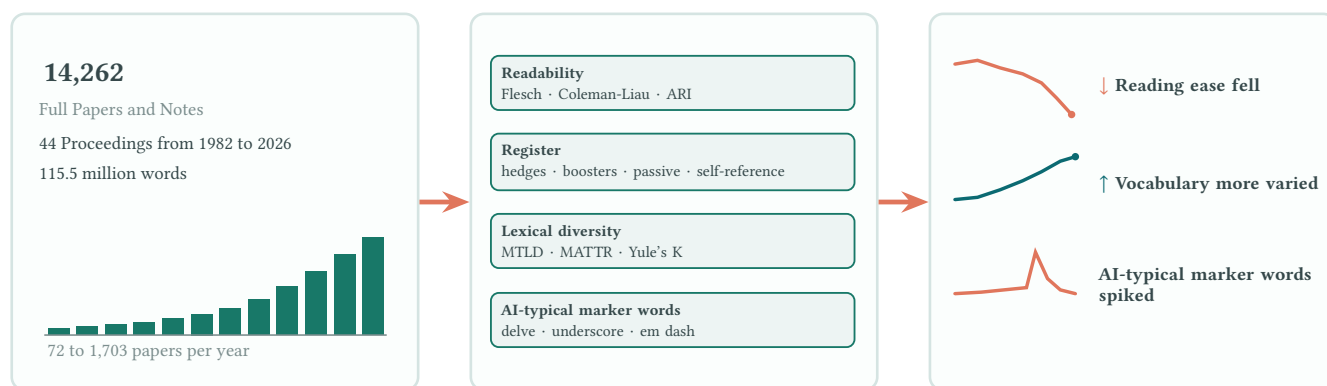
\begin{teaserfigure}
  \centering
  \resizebox{\textwidth}{!}{
    % Three-panel teaser for "How Did Writing Change At CHI?"
% Requires: tikz with libraries calc, positioning, arrows.meta
% Requires colors: deepteal, wordteal, coral, ink, muted, hair, cardbg
\begin{tikzpicture}[
  font=\small,
  >=Stealth,
  panel/.style={draw=hair, line width=1pt, rounded corners=5pt, fill=cardbg},
  ptitle/.style={font=\bfseries, text=deepteal},
  bignum/.style={font=\bfseries\Large, text=ink},
  lbl/.style={text=ink, inner sep=1pt},
  mut/.style={text=muted, font=\footnotesize, inner sep=1pt},
  chip/.style={draw=wordteal, line width=0.7pt, rounded corners=3pt, fill=wordteal!8,
               text=ink, align=left, inner sep=4pt, font=\scriptsize,
               text width=4.55cm, minimum height=0.72cm},
  flow/.style={-{Stealth[length=3.6mm,width=2.8mm]}, line width=1.8pt, coral},
]

% ---------- panels ----------
\node[panel, minimum width=5.4cm, minimum height=5.0cm, anchor=south west] (A) at (0,0){};
\node[panel, minimum width=5.4cm, minimum height=5.0cm, anchor=south west] (B) at (6.1,0){};
\node[panel, minimum width=5.4cm, minimum height=5.0cm, anchor=south west] (C) at (12.2,0){};

% ---------- flow arrows ----------
\draw[flow] (A.east) -- (B.west);
\draw[flow] (B.east) -- (C.west);

% ================= Panel A : the corpus =================
\node[bignum, anchor=north west] at ([shift={(0.32,-0.55)}]A.north west) {14{,}262};
\node[mut, anchor=north west] at ([shift={(0.34,-1.15)}]A.north west) {Full Papers and Notes};
\node[lbl, anchor=north west, font=\footnotesize] at ([shift={(0.34,-1.6)}]A.north west) {44 Proceedings from 1982 to 2026};
\node[lbl, anchor=north west, font=\footnotesize] at ([shift={(0.34,-2.0)}]A.north west) {115.5 million words};

% growth bars (schematic: papers per year, 72 to 1,703)
\foreach [count=\i from 0] \h in
  {0.09,0.11,0.14,0.17,0.22,0.28,0.35,0.47,0.64,0.84,1.07,1.30}{
  \fill[wordteal] ([shift={(0.5+\i*0.38,0.75)}]A.south west) rectangle ++(0.28,\h);
}
\draw[muted, line width=0.6pt] ([shift={(0.45,0.75)}]A.south west) -- ([shift={(5.0,0.75)}]A.south west);
\node[mut, anchor=north west] at ([shift={(0.45,0.67)}]A.south west) {72 to 1{,}703 papers per year};

% ================= Panel B : what we measured =================

\coordinate (Bc) at ([shift={(2.7,0)}]B.south west); % centre x of panel B is x of Bc

\node[chip] (c1) at (Bc |- 0,4.10) {\textbf{Readability}\\[1pt] Flesch $\cdot$ Coleman-Liau $\cdot$ ARI};
\node[chip] (c2) at (Bc |- 0,3.05) {\textbf{Register}\\[1pt] hedges $\cdot$ boosters $\cdot$ passive $\cdot$ self-reference};
\node[chip] (c3) at (Bc |- 0,2.00) {\textbf{Lexical diversity}\\[1pt] MTLD $\cdot$ MATTR $\cdot$ Yule's K};
\node[chip] (c4) at (Bc |- 0,0.95) {\textbf{AI-typical marker words}\\[1pt] delve $\cdot$ underscore $\cdot$ em dash};

% ================= Panel C : what we found =================

% Row 1 : reading ease down
\draw[coral, line width=1.3pt, line cap=round]
  plot coordinates {(12.55,4.35)(12.85,4.4)(13.15,4.3)(13.45,4.22)(13.7,4.1)(13.9,3.9)(14.1,3.68)};
\fill[coral] (14.1,3.68) circle (1.4pt);
\node[anchor=west, text width=2.95cm, font=\footnotesize, text=ink] at (14.35,4.05)
     {\textcolor{coral}{$\downarrow$}\ \textbf{Reading ease fell}};

% Row 2 : diversity up
\draw[deepteal, line width=1.3pt, line cap=round]
  plot coordinates {(12.55,2.55)(12.85,2.58)(13.15,2.68)(13.45,2.8)(13.7,2.92)(13.95,3.06)(14.15,3.12)};
\fill[deepteal] (14.15,3.12) circle (1.4pt);
\node[anchor=west, text width=2.95cm, font=\footnotesize, text=ink] at (14.35,2.75)
     {\textcolor{deepteal}{$\uparrow$}\ \textbf{Vocabulary more varied}};

% Row 3 : marker words spike then fade
\draw[coral, line width=1.3pt, line cap=round]
  plot coordinates {(12.55,1.3)(12.9,1.32)(13.2,1.35)(13.5,1.38)(13.62,1.85)(13.78,1.5)(13.95,1.35)(14.15,1.3)};
\node[anchor=west, text width=2.95cm, font=\footnotesize, text=ink] at (14.35,1.4)
     {\textbf{AI-typical marker words spiked}};

\end{tikzpicture}
  }
  \caption{Overview of the study. We analyzed 44 CHI proceedings, including 14,262 archival papers, and 115.5 million words from 1982 to 2026. Every paper passed through a single pipeline that assessed readability, register, lexical diversity, and the presence of published AI-typical marker words, with function-word and same-author controls. Word complexity rose, and reading ease fell, crossing into the band where the Flesch scale is very difficult in 2022. Yet vocabulary grew more varied rather than less, which runs against the claim of homogenization, and AI-typical marker words spiked around 2024 before fading. The change is stylistic, and its onset predates the first cohort of authors who could have used these tools. We therefore read generative AI as an accelerant of a long drift toward density rather than its cause.}
  \Description{A three-panel overview joined by arrows, above a summary banner. Panel one, the corpus, reports 14,262 archival papers across 44 editions from 1982 to 2026 and 115.5 million words, with a small bar chart of the venue growing from 72 to 1,703 papers per year. Panel two, what we measured, shows the full-body text feeding into four measure groups: readability, register, lexical diversity, and marker words, with function-word and same-author controls. Panel three, what we found, shows three small trend lines: reading ease falling into the very difficult band by 2022, vocabulary growing more varied rather than less against homogenization, and marker words spiking then fading with the onset before LLMs, noting that 20 of 25 measures record their largest three-year shift on record. The banner reads that prose densified rather than homogenized and that generative AI accelerated a four-decade drift rather than starting it.}
  \label{fig:teaser}
\end{teaserfigure}

\maketitle

\section{Introduction}

% Context
Large Language Models (LLMs) have entered academic workflows at scale. Since writing has always drifted through all times in grammar and in lexical fashion with every technology released, general access to LLMs changed how fast and in which direction scholarly prose moves~\cite{waltherDutordoir2026, comasforgas2026, comasforgas2025}. Drafting, revising, and paraphrasing with an LLM has become commonplace~\cite{hutson2022, salvagno2023, kousha2024}. Consequently, a growing body of work documents a matching shift in the published record itself. Studies of abstracts and titles report a sharp post-2022 rise in a set of AI-typical marker words such as \emph{delve}, \emph{underscore}, and \emph{intricate}~\cite{comasforgas2026, kobak2025, matsui2025, uribe2024, comasforgas2025}. These results are usually read as evidence that models are narrowing scientific language and pushing prose toward a single generic lexical register.

% Why HCI, and why CHI in particular
Human-Computer Interaction (HCI) has no reason to be exempt from this trend, and the CHI conference, the largest HCI conference to date, is experiencing an increase in submissions, especially since the advent of publicly available LLMs. Prose is part of what CHI reviews, and as such, it is relevant to study how LLMs affected the written prose of CHI papers, as it can be decisive for paper decisions. Reviewers weigh clarity, framing, and argument alongside the contribution, so a drift in what reads as normal academic writing moves the target that authors and reviewers calibrate against~\cite{jung2026drivespaperacceptanceprocesscentric}. CHI published 44 proceedings, providing over four decades of continuous linguistic practice against which changes in prose can be measured. If that practice is moving, and if linguistic features carry weight in the review and publication process, then the use of tools such as LLMs in writing affects or even shapes the community. Understanding how CHI papers have changed linguistically in response to LLMs is, therefore, a prerequisite for reasoning about and assessing paper quality. 

% Research gap
Prior work investigated how writing has changed across a range of academic fields since the rise of LLMs, including research papers in science~\cite{liang2024monitoring}, finance~\cite{waltherDutordoir2026}, dental research~\cite{uribe2024}, and biomedical research~\cite{kobak2025, matsui2025}. However, previous studies are affected by limitations, such as analyzing abstracts or titles, the shortest and most heavily edited part of a paper~\cite{comasforgas2026, kobak2025, matsui2025}, rather than the full corpus of a proceedings series. Furthermore, prior work considered a fixed list of AI-typical marker words, which rises and falls with topic and fashion as much as with writing style, and none of the published lists has been shown to transfer beyond the literature from which it was derived~\cite{comasforgas2026, comasforgas2025, waltherDutordoir2026}. Finally, prior work rarely compared its results to a baseline from the pre-LLM era. The claim that LLMs homogenize writing~\cite{sourati_shrinking_2026} has not been tested across an entire series of scientific proceedings, and its support largely comes from short controlled writing tasks rather than from a published record observed over many years~\cite{comasforgas2026, padmakumar2024, doshi2024, anderson2024}.

% How we address the gap
We address these gaps with the full-body text of every paper contribution to CHI from 1982 to 2026, totaling 14,262 papers and Notes\footnote{CHI Notes were the short paper track that existed between CHI 2006 and 2017.} (see \autoref{fig:teaser}). In addition to AI-typical marker words~\cite{comasforgas2026, comasforgas2025, waltherDutordoir2026}, we compute readability, register, and diversity measures that do not rely on a fixed word list. Furthermore, we measured both across documents and within them, comparing the phrasing papers share and the distance between whole yearly style profiles. 
% Most important findings
Our findings show that word-level complexity increased, while reading ease decreased. The estimated break falls in the early 2020s rather than in 2024, when LLMs became popular, placing their onset at or before the first cohort that could have used these LLMs. Against this, we could not find evidence that CHI prose became more homogenized, as is the case in other domains~\cite{sourati_shrinking_2026}. Vocabulary became more varied rather than less on size-matched diversity measures, and sentence rhythm did not become more uniform. Shared phrasing across papers, the one measure that tests convergence directly, did rise, but it began rising in 2021 and returned to its historical level in 2026. Function words alone predict whether a paper is recent, so the change is stylistic and not merely a change of subject. Yet the whole function-word profile barely moved, which places the shift in a few dimensions rather than in a broad restructuring of style.

\aptLtoX{\begin{framed}
\noindent\textbf{Contribution Statement:} Our contribution is fourfold. (1) We provide the first full-text analysis of CHI writing across the LLM transition. It covers all 44 proceedings and the full text of every paper, and is measured against the full history of CHI. (2) We show that CHI prose grew denser and words got longer while reading ease fell, starting before the rise of LLMs. (3) We test the claim that AI makes writing more uniform, finding that vocabulary grew more varied. (4) Finally, we test published AI-typical marker words on CHI.
\end{framed}}
{\begin{flushleft}
\setlength{\fboxsep}{10pt}%box to content distance
\setlength{\fboxrule}{2pt}%thickness of box
\fcolorbox{gray!60}{white}{%
    \parbox{.89\columnwidth}{%
    \textbf{Contribution Statement:} Our contribution is fourfold. (1) We provide the first full-text analysis of CHI writing across the LLM transition. It covers all 44 proceedings and the full text of every paper, and is measured against the full history of CHI. (2) We show that CHI prose grew denser and words got longer while reading ease fell, starting before the rise of LLMs. (3) We test the claim that AI makes writing more uniform, finding that vocabulary grew more varied. (4) Finally, we test published AI-typical marker words on CHI.
    }
}
\end{flushleft}}

\section{Related Work}
The following section presents prior work. We begin with how LLMs have been used to support academic writing. We then show how lexical traces LLMs have left in scientific text since they became widely available. We present work that investigated the homogenization of scientific writing. Finally, we close with the measures of register and lexical diversity that inform our analysis. We then present our research questions.

\subsection{LLM Support for Writing and Research}
HCI has studied AI writing tools for years to support users in writing tasks. Lee et al. built CoAuthor, a large dataset of human-AI writing sessions, to probe what language models can and cannot do as writing partners~\cite{lee2022coauthor}, with a later survey of the field of AI-assisted writing~\cite{lee2024designspace}. As a result, the authors lay out a design space across task, user, technology, interaction, and ecosystem. Jakesch et al. hypothesized that AI-assisted writing can change the writing style of the writer~\cite{jakesch2023}. The authors gave 1,506 participants a writing assistant that leaned toward one side of a debate, and the assistant moved what people wrote and what they later reported believing. Thus, LLM-assisted writing may change opinions about statements and written text, potentially shifting the ownership of who owns LLM-generated text. In this context, Draxler et al. report an AI-Ghostwriter effect on authorship, where writers declared themselves authors of text they did not feel they owned~\cite{draxler2024}. 

If co-writing moves opinions and blurs ownership, it can also move writing styles, which our work measures at scale. In this context, generative AI has also become a tool for supporting HCI research. On the one hand, LLMs are now used to annotate text at scale~\cite{gilardi2023} and to generate synthetic user study data~\cite{hamalainen2023}, both of which assume that the text they consume and produce is stable and human-like. On the other hand, the use of LLMs to support scientific research has been criticized for several reasons, including misrepresenting diverse voices~\cite{10.1145/3613904.3642703}, undermining reproducibility~\cite{10.1145/3695765}, and potentially manipulating research results through strategic prompting~\cite{10.1145/3744911}. However, a drifting linguistic baseline complicates that assumption. The same tools have entered scholarly evaluation, where Liang et al. estimated that a measurable share of recent conference peer reviews was modified by LLMs~\cite{liang2024monitoring}. AI is now part of how HCI writes, studies text, and reviews. In contrast, prior work examined the writing process, single artifacts, or controlled sessions, and no prior work has investigated how writing at CHI has changed before and after the LLM era. We close this gap by investigating how writing and its lexical features have changed at CHI since the widespread introduction of LLMs.

\subsection{Measuring Registers}
Our measures build on established work in corpus linguistics and readability, which are referred to as registers. Change in academic writing predates LLMs and has accompanied technological developments, such as the proliferation of the internet and search engines. For example, Plav{\'e}n-Sigray et al. stated that readability of scientific texts has declined for over a century~\cite{plavensigray2017}, and academic prose has slowly shifted its register, for example, toward more informal features in some fields~\cite{hylandjiang2017}. Readability has also been reported to fall further after the release of ChatGPT~\cite{alsudais_exploring_2025}, though the formulas involved were calibrated on school prose and are not comparable to scientific texts. Any claim about a recent break has to be read against this long drift, which is why we benchmark our measures against CHI's history. Furthermore, the writing of a CHI paper can change more than its vocabulary\footnote{For example, CHI submission format, page limits, and paper tracks changed over the years.}, and research in NLP has traced how the types of contributions papers make have shifted over time~\cite{pramanick_nature_2025}. Thus, the scope of our paper is the lexical development of writing.

We group measures into three families. Readability formulas summarize how demanding a text is to read. We use Flesch reading ease~\cite{flesch1948} together with the character-based Coleman-Liau index~\cite{colemanliau1975} and the Automated Readability Index~\cite{ari1967}, which avoids syllable counting. We measure register through hedges, boosters, and self-reference, which mark the writer's stance in academic prose~\cite{hyland2018metadiscourse, biber1988}. Lexical diversity has a long history of length-sensitive measures, where we measure type-token ratio and Yule's K~\cite{yule1944} depend on text length, so we also use Moving-Average Type-Token Ratio (MATTR)~\cite{mattr2010} and Measure of Textual Lexical Diversity (MTLD)~\cite{mtld2010}, which were designed to stay stable across lengths, and Shannon entropy~\cite{shannon1948}, at the corpus level.

\subsection{Homogenization of Writing and Lexical Diversity}
Previous work states that LLM usage makes writing more homogeneous. Comas-Forgas et al. stated that AI pushed scholarly language toward a polished, generic register and warned that individual writing styles are being eroded~\cite{comasforgas2026}. Controlled studies provide partial support for the claim. Padmakumar and He show that co-writing with an instruction-tuned model reduces the diversity of argumentative essays across authors~\cite{padmakumar2024}. Doshi and Hauser find that access to model-generated ideas raises the quality of individual stories but makes them more similar to one another~\cite{doshi2024}. Anderson et al. report the same convergence for ideation, where users of a language model align at the group level even as each feels more productive~\cite{anderson2024}. The shift is toward greater informality, with human and AI abstracts compared against ground-truth labels and higher scores reported for AI-associated text on features such as first-person pronouns and listing expressions~\cite{zhao_informality_2026}. Informality is a single evaluation factor, and the field does not agree on its components, so we decompose it into the features of self-mention and passive voice. We report each measure on its own. Finally, Sourati et al.~\cite{sourati_shrinking_2026} analyzed how LLMs affect writing on 880,000 text records. Their results show that, while the core content remains the same, texts become homogenized and reduce writing complexity. However, the authors' work examined a diverse set of text records across different fields. In contrast, our work investigates CHI as a single venue containing scientific texts.

\subsection{Lexical Traces of LLMs}
A growing body of literature measures how the vocabulary of scientific writing has changed since the release of ChatGPT. Previous research shows that specific words have become far more frequent after 2022. Kobak et al. estimated excess word usage across millions of biomedical abstracts and found that a small set of marker words rose sharply in 2023 and 2024~\cite{kobak2025}. The same pattern appears more broadly across other domains, such as PubMed~\cite{matsui2025} and dental research~\cite{uribe2024}. Comas-Forgas et al. moved from single words to a curated set of markers and identified 17 lexical items with steep post-2022 growth in Scopus abstracts~\cite{comasforgas2026}. Related work reports a parallel rise in the use of participial verbs in article titles~\cite{comasforgas2025}. Similar findings were found by other researchers~\cite{waltherDutordoir2026, yakura_empirical_2026}. Across these studies, words such as ``delve, ``underscore,'' and ``intricate were widely read as signs of AI-assisted prose.

However, the AI-typical marker words themselves are not stable across time. Geng and Trotta tracked AI-typical marker words over time and reported that several AI-typical marker words dropped sharply once it was clear that they were more likely to be produced by an LLM in early 2024, while others, such as significant, kept rising~\cite{geng_human-llm_2025}. A list validated for one year, therefore, has a short shelf life, which is directly applicable to any proposal to use such lists to detect machine-assisted writing. Second, the shift is not confined to text. Yakura et al. found the same words rising in spontaneous academic speech on podcasts, causally linked to the release of ChatGPT through a synthetic control condition~\cite{yakura_empirical_2026}. If the vocabulary has entered speech, a paper need not quote a model to carry it, which matters for reading any body-text rise as a change in writing rather than in subject matter.

Furthermore, prior research investigated the extent to which whole paper bodies were modified by LLMs. Liang et al. developed a population-level estimator and applied it to more than a million papers, reporting the fastest growth in computer science, where up to 22\% of papers show signs of LLM modification by late 2024~\cite{liang2024mapping}. The authors also found that modifications were higher among first authors who post preprints often and in more crowded research areas. Metadata studies add that explicit acknowledgment of AI assistance remains rare, so observed lexical change far outruns actual disclosed use~\cite{kousha2024}.

Prior work is affected by several limitations. First, previous work measured abstracts or titles rather than the body of the paper~\cite{comasforgas2026, comasforgas2025, kobak2025, matsui2025}. The one exception at scale is Kousha and Thelwall, who track twelve marker terms across 2.4 million biomedical open-access full texts and report the same steep post-2022 rise in the body~\cite{kousha2026much}. Their corpus is biomedical, and their instrument is a fixed-term list. However, we focus on CHI across all 44 existing volumes and compare it with its own history. Furthermore, we add AI-typical marker words with register and diversity measures to understand how AI-typical marker words changed over time. Second, marker-word designs track topic and fashion as much as writing style, and adoption varies with author attributes such as discipline and first-language status~\cite{lin2025divergentllmadoptionheterogeneous}. A raw lexical shift, therefore, mixes a change in writing with a change in who writes. Our function-word and same-author controls are designed to distinguish between the two.

\subsection{Detecting LLM-Written Text}
Previous research has increasingly focused on the automatic detection of LLM-written text. For example, zero-shot detectors read statistical signatures of machine text, such as entropy, rare n-gram frequency, perplexity, and token log-probabilities~\cite{lavergne2008, badaskar2008, beresneva2016, solaiman2019}. DetectGPT scores the log-probability curvature of a passage~\cite{mitchell2023detectgpt}, DNA-GPT compares n-gram divergence~\cite{yang2023dnagpt}, Fast-DetectGPT swaps in conditional probability curvature for speed~\cite{bao2023fastdetectgpt}, and various other methods estimate the intrinsic dimension of the text~\cite{tulchinskii2023}. A second family of classifiers on labeled human and machine text~\cite{10.1162/COLIa00166, zellers2019, bakhtin2019, uchendu2020}. These finetune language-model backbones for the binary decision~\cite{chen2023gptsentinel, yu2023gptpat, li2023deepfake} and add contrastive or adversarial objectives to make the classification more robust~\cite{liu2022coco, bhattacharjee2023conda, hu2023radar}.

However, LLM-detection approaches have severe limitations. Zero-shot methods need access to model internals, and a closed model forces the use of a proxy that weakens the signal. Trained classifiers overfit the models they learned from, break under simple adversarial edits~\cite{wolff2020}, and provide false positives on non-dominant language varieties (i.e., when writing text that does not follow common writing standards~\cite{liang2023biased}. The reliability of public detectors has been questioned repeatedly~\cite{openai2019gpt2, jawahar2020, fagni2021, ippolito2019, gehrmann2019gltr, heikkila2022, crothers2022}, and OpenAI withdrew its own classifier for low accuracy~\cite{kirchner2023, kelly2023}. Whether accurate instance-level detection is possible at all is still contested~\cite{weberwulff2023, sadasivan2023, chakraborty2023}. We therefore measure lexical change across the whole corpus, where population-level estimates remain stable even when single-document calls are not~\cite{liang2024monitoring}.

\subsection{Summary and Research Questions}
The academic community now writes with LLMs, studies text with them, and reviews with them, yet the field has examined writing sessions and single artifacts rather than its own published record~\cite{lee2022coauthor, jakesch2023, draxler2024, liang2024monitoring}. The lexical-trace literature in other fields documents a sharp post-2022 shift, but it measures abstracts and titles and relies on fixed marker lists that follow topic and fashion as closely as style~\cite{kobak2025, comasforgas2026, comasforgas2025}. Researchers are afraid that the homogenization of scientific text reduces lexical diversity. Yet, this claim is the least tested, and its support comes from short, controlled tasks and single sessions~\cite{padmakumar2024, doshi2024, anderson2024}, not from a large published record at venues such as CHI, as has been observed over many years. Prose is part of what CHI reviews, so a change in what counts as rigorous shifts the bar that authors and reviewers calibrate against, and that bar determines what gets published and by whom. Newcomers and authors who do not write in English as a first language carry most of that cost. CHI is also one of the few communities positioned to respond, since it writes its own review guidance and holds four decades of its own record to measure against. Related work revealed three gaps. (1) Full paper bodies are largely unmeasured, (2) marker-word designs cannot separate a change in style from a change in subject, and (3) the claim that AI narrows academic prose has not been tested where a long publication history could serve as its own baseline. To investigate how writing at CHI has changed, we analyzed all 44 existing volumes of CHI and compared the shift in writing. Size-matched diversity and register measures let us ask whether vocabulary narrowed or widened. Function-word and same-author controls let us separate how CHI writes from what it writes about and who writes it. We investigate the following three research questions.
 
\begin{description}
  \item[\textbf{RQ1:}] How have the lexical properties of CHI writing changed over the history of the conference since LLM writing tools became available?
  \item[\textbf{RQ2:}] How does the change since LLM writing tools became available compare to previous volumes of CHI editions?
  \item[\textbf{RQ3:}] To what extent does the recent shift show homogenization toward a narrower register?
\end{description}

\section{Dataset}
\label{sec-dataset}
Our analysis draws on a corpus of 14,262 archival contributions published at the ACM CHI Conference on Human Factors in Computing Systems between 1982 and 2026. The corpus covers all 44 proceedings of the conference. We included all CHI papers from 1982\footnote{CHI did not take place in 1984.} in our analysis. The corpus includes 13{,}340 full papers and 845 Notes. Notes was a four-page archival short-paper track at CHI, which ran from 2006 to 2017. 77 archived items are between four and six pages and were published between 2016 and 2017, where references did not count towards the Notes page limit. 
We assembled the corpus from the ACM Digital Library, which has been fully open access since 2026. Full papers and Notes published between 2006 and 2026 were labeled as papers and obtained from the ACM Digital Library. The proceedings published between 1982 and 2005 did not label papers' metadata and instead grouped them into different tracks. We describe the paper history and paper classification process below.

\begin{figure*}[t]
  \centering
  \begin{minipage}[t]{0.56\linewidth}
    \vspace{0pt}
    \centering
    \includegraphics[width=\linewidth]{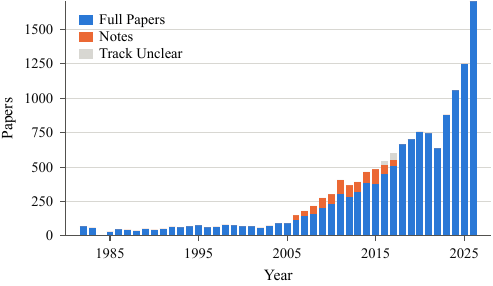}
    \caption{Corpus size and track composition by year used for analysis. Notes were introduced in 2006 and discontinued after 2017. CHI grew 24-fold across the period, which is why all later measures are reported as rates and, where they depend on length, size-matched. The track is unclear for 77 items as ACM did not label the papers as full papers or Notes. Since the 77 submissions have four pages of content and references, we suspect that these are Notes where references did not count towards the page limit in 2016 and 2017 after manual inspection.}
    \Description{A stacked bar chart of papers per year from 1982 to 2026. Full papers we could obtain rose from 72 in 1982 to 1703 in 2026. An orange Notes segment appears from 2006, peaks at around 100 papers per year, and disappears after 2017. A small grey segment in 2016 and 2017 marks 77 items of six or seven pages, a length that identified neither track once the page limits changed. Six papers could not be obtained.}
    \label{fig-corpus}
  \end{minipage}\hfill
  \begin{minipage}[t]{0.40\linewidth}
    \vspace{0pt}
    \centering
    \captionsetup{type=table}
    \caption{CHI archival contributions per conference year. CHI was not held in 1984. The numbers indicate the records we obtained that qualify as archived paper. All contributions through 2005 were labeled as papers, regardless of whether they were archived paper contributions or, for example, panels. Such submissions were not included in the analysis.}
    \label{tab-corpus}
    \small
    \begin{tabular}{@{}rr@{\hspace{2em}}rr@{\hspace{2em}}rr@{}}
      \toprule
      Year & $n$ & Year & $n$ & Year & $n$ \\
      \midrule
      1982 & 72 & 1997 & 67 & 2012 & 369 \\
      1983 & 58 & 1998 & 81 & 2013 & 392 \\
      1985 & 29 & 1999 & 78 & 2014 & 465 \\
      1986 & 48 & 2000 & 72 & 2015 & 484 \\
      1987 & 46 & 2001 & 69 & 2016 & 545 \\
      1988 & 39 & 2002 & 61 & 2017 & 600 \\
      1989 & 54 & 2003 & 75 & 2018 & 665 \\
      1990 & 46 & 2004 & 93 & 2019 & 702 \\
      1991 & 52 & 2005 & 93 & 2020 & 758 \\
      1992 & 67 & 2006 & 151 & 2021 & 746 \\
      1993 & 63 & 2007 & 182 & 2022 & 637 \\
      1994 & 69 & 2008 & 218 & 2023 & 879 \\
      1995 & 76 & 2009 & 277 & 2024 & 1{,}057 \\
      1996 & 64 & 2010 & 302 & 2025 & 1{,}249 \\
      \multicolumn{2}{c}{} & 2011 & 409 & 2026 & 1{,}703 \\
      \midrule
      \multicolumn{5}{r}{Total} & 14{,}262 \\
      \bottomrule
    \end{tabular}
  \end{minipage}
\end{figure*}

\subsection{The History of the CHI Paper Track}
\label{sec-track-history}
The paper track of CHI has changed several times, which is relevant for describing our dataset. From 1982 to 1993, the main proceedings accepted only full papers, with no archival short-paper track (i.e., Notes). The first page limit in the corpus was eight pages, enforced from 1994 to 2004, and was raised to ten pages from 2005 to 2015. Until 2015, both limits counted the reference list, and references did not count towards the page length after 2015. CHI introduced the Notes track in 2006, a four-page archival short paper that also counted references, being in place through 2017. In 2016 and 2017, references no longer counted toward the page limit. Notes were discontinued in 2018, leaving only full papers. In 2021, CHI abolished page limits and moved to a word limit.

\subsection{Identifying Papers From 1982 to 2005}
\label{sec-dataset-early}
The proceedings from 1982 to 2005 require special consideration because they are not labeled as full papers in the ACM metadata. We describe our retrieval process below.

\subsubsection{Volume Identification}
We enumerated the 23 CHI \emph{main} proceedings volumes for 1982 to 2005 from the ACM conference listing. For each volume, we extracted the complete table of contents to obtain a list of all papers, comprising DOI, title, authors, page range, and item type. Working from the main proceedings volume alone means that the separately published ``Extended Abstracts'' and ``Conference Companion'' volumes were excluded, as they include posters, demonstrations, doctoral consortium contributions, and SIG meetings. Two volumes are joint conferences and were retained whole, as ACM publishes them. These are CHI 1987, published with the Graphics Interface conference, and CHI 1993, published with the INTERACT conference.

\subsubsection{Rule-Based Filtering}
From the 1{,}714 items collected, we removed 206 that were not archival papers. These include front matter and indexes (2 items), items whose ACM type is not an article (1 item), items carrying an explicit track marker in the title such as \emph{(panel session)}, \emph{(lab review)}, or \emph{(plenary address)} (43 items), and items falling below the minimum page lengths for CHI papers (157 items). We modeled this exclusion, aware of the CHI edition, because the page conventions have changed over time. For example, a four-page CHI 1982 item was a refereed paper, whereas by the 1990s, the two- to three-page submissions consisted almost entirely of panels, laboratory reviews, and video or demonstration abstracts. We manually inspected the excluded items to avoid false positive exclusions.

\paragraph{Resolving Ambiguous Items From Document Content}
Through CHI 1993, the main proceedings integrated papers with panels and laboratory reviews of comparable length. Neither page count nor ACM metadata separates them. We manually classified these papers from the text of the PDFs themselves. Panel and laboratory-review summaries identify their format on the first page, using markers such as \emph{Panelists}, \emph{Moderator}, \emph{Panel organizers}, and \emph{Panel organizers}, or a running header such as \emph{INTERCHI 1993 Panel}. Refereed papers carry no such marker or headline. Thus, we manually inspected these papers and removed those that pointed to panels.

The content-based step materially changed the corpus relative to a title-based reading. Thus, through manual inspection, we retained short refereed papers that a categorical heuristic would discard (e.g., papers that would not be classified as papers although they are papers based on markers), including seminal papers of Lee et al. ``A multi-touch three-dimensional touch-sensitive tablet'' at CHI 1985~\cite{10.1145/317456.317461} (five pages long) and the ``The perspective wall'' at CHI 1991 by Mackinlay et al.~\cite{mackinlay1991perspective} (seven pages). Our inspection also removed rhetorically titled multi-author items that are in fact panels. For example, the two papers ``The interface is often not the problem'' at CHI 1986~\cite{10.1145/1165387.30872} and ``The art of the obvious'' at CHI 1992~\cite{10.1145/142750.142800} read as panel titles but are single-institution refereed papers. In contrast, ``Should we or shouldn't we use spoken commands in voice interfaces?'' at CHI 1991~\cite{brennan1991should} is a panel.

\subsection{Identifying Papers From 2006 to 2026}
\label{sec-dataset-recent}
From 2006 onward, the ACM Digital Library labels each contribution by type, so full papers and Notes are read directly from the metadata. We retained every item labeled as a full paper or note in the main proceedings volume of each year. The separately published Extended Abstracts and Companion volumes were excluded. Notes are labeled through 2017, the last year the track ran.

\subsection{Corpus Composition}
Of the 1,478 full papers identified from 1982 to 2005, we obtained 1,472, with 6 missing. The six exceptions have no full text in the ACM Digital Library at all, since their article pages resolve but offer no PDF. We confirmed each one individually. \autoref{tab-corpus} reports the corpus by year and~\autoref{fig-corpus} shows how its composition changes. Our dataset shows that CHI grew from 72 accepted papers in 1982 to 1,703 in 2026. The displayed numbers are papers we could obtain as a PDF file through the ACM Digital Library, excluding those that did not offer a PDF\footnote{For example, retracted papers were unavailable.}. 

\subsection{Track Composition}
\label{sec-tracks}
\begin{figure}[t]
  \centering
  \includegraphics[width=\columnwidth]{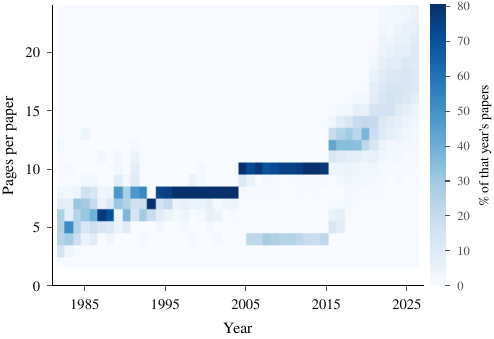}
  \caption{Distribution of paper length by year, as a share of each year's papers. Notably, two bands are visible, reflecting the eight-page limit between 1994 and 2004 and the ten-page limit between 2005 and 2015, both of which include references. Furthermore, CHI held a Notes track with four pages of references between 2006 and 2015. However, the conference dropped the reference to this page limit in 2016 and 2017. Thus, from 2016 onward, the distribution fills in, and the length no longer identifies the track, which could be either a paper or a Note. Notes were removed from 2018 onward. Page limits were removed in 2021 and replaced with a word limit.}
  \Description{A heatmap with year on the horizontal axis and pages per paper on the vertical axis, shaded by the percentage of that year's papers. A dark band is at ten pages and a lighter band at four pages from 2006 to 2015, with no papers at all between six and seven pages. After 2015, the distribution spans 11-14 pages.}
  \label{fig-tracks}
\end{figure}

The CHI submission tracks changed twice during the period (see~\autoref{fig-tracks}). We recovered the composition from document length, which provides an initial classification of the track. Papers and Notes had fixed and non-overlapping page limits. In every year from 2006 to 2015, the corpus contains a cluster of four to five pages and a cluster of ten pages.

The years from 1982 to 2005 contain only full papers because CHI had no archival short-paper track in the main proceedings, and the median contribution length runs eight pages from 1994 onward. Notes appear in 2006 and account for between 18\% and 28\% of the proceedings annually through 2015. Their share fell to 12\% in 2016 and 8\% in 2017, before the track was discontinued. From 2018, the distribution is unimodal, and paper length grows steadily, with the median rising from 12 pages in 2017 to 20 pages in 2026. Each item is labeled by track in the released index, so the corpus can be restricted to full papers alone.

\subsection{Verification}
All files obtained were valid PDFs, and no DOI appears more than once. Every filename we obtained encodes the year and DOI of the record it corresponds to. Every item belongs to the main proceedings volume of its year, so no Extended Abstracts entered the corpus. Our manual check also removed 50 files that were not PDFs at all, but HTML landing pages saved with a .pdf extension by an earlier retrieval pass, produced when ACM holds no full text for a DOI. These 50 items were non-archival, such as session dividers, conference welcomes, ``CHI Madness'' entries, and panels or SIGs without full text. No paper was lost, but these files would otherwise have been treated as silent noise in a text-extraction pipeline.

For the years from 1982 to 2005, we additionally compared each page count against the page range recorded by ACM. Of 1,472 items, 1,411 agree exactly. Manual inspection of the page discrepancies showed they were misclassifications in the page metadata of ACM rather than truncated retrievals. Positive page discrepancies correspond to color-plate pages appended to the scans. Analyses that depend on paper length should therefore use the page count of the document rather than the page range in the bibliographic record.

\section{Method}
\label{sec-method}
All analyses were run over the entire previously described CHI corpus, including 14,262 archival contributions and 115.5 million tokens of body text. A token represents one word in our analysis.

\subsection{Text Preparation}
\label{sec-method-prep}
The body text, abstract, acknowledgments, and reference list were extracted from each PDF using PyMuPDF\footnote{\url{https://pymupdf.readthedocs.io/en/latest}}. Every CHI proceeding is set in two columns, so text blocks are explicitly ordered by column rather than by their order in the PDF content stream, which is unreliable for early-scanned volumes. Full-width elements, such as titles and wide figures, act as band separators so they do not scramble the columns around them. Running heads, bare page numbers, and figure captions are removed, line-break hyphenation is rejoined, and the body is anchored at the ``Introduction'' heading so that publisher boilerplate does not enter the text\footnote{For example, the CCS concepts and paper keywords.}.

\subsubsection{Removing Publisher Boilerplate Text}
Anchoring the body at the ``Introduction'' heading removes the front matter but not the ACM permission notice, which the publisher prints in the left column of the first page and which therefore falls inside the body. The notice is identical across papers, so any measure of shared phrasing counts it as writing. Its incidence is also strongly year-dependent, which is worse than a constant contamination because it manufactures trends where none exist. Running headers behave the same way. Left in place, this alone produced an apparent decline in templated phrasing across the period of interest, which disappeared from the literature corpus once we removed it.

We removed the publisher boilerplate text in four passes. The notice itself is found on the token stream rather than on lines, because optical character recognition in the scanned volumes breaks it into one word per line and corrupts the words. A window of 22 tokens is scored by the share of its content words drawn from the canonical notice, and a window at or above 0.70 with at least six distinct matches is cut. On 1,603 sampled documents, this found 800 spans, of which one was not a notice. The tail lines, meaning the conference line, the copyright line, the ISBN, the DOI, and the Creative Commons sentence, carry no notice vocabulary and are found instead by growing outward from lines that only the block contains. Isolated fragments are removed by exact phrase, since the extractor sometimes strands one line of the notice between two paragraphs of the introduction, and in CHI 2020, one such line gets through the first pass in 492 of 758 papers. Running headers are removed as any line that repeats three or more times in one document, a rule that names no venue and thus catches the 2000 imprint line and the spaced-out 1985 page header alike. The correction removes 1.09\% of corpus tokens, ranging from 0.26\% in 2019 to 3.81\% in 1985.

% [MA] "How do you define a sentence? Does a ':' end a sentence?"
% [MA] "I guess defined as number of words/tokens? Again, should be specified."
\subsubsection{Tokens, Sentences, and Normalization}
The tokenizer matches a letter followed by letters, apostrophes, or hyphens, so punctuation never becomes a token, and the two terms are interchangeable throughout this paper. For example, \emph{Tom's} is one token and \emph{human-computer} is one token because a tokenizer that splits either would change every rate we report. Text is lowercased before counting, so \emph{The} and \emph{the} are the same type. We do not lemmatize anywhere except in the AI-marker replication, where the published lists are defined in terms of stems, and matching them requires lemmatization. Stemming is avoided elsewhere because it would merge inflections whose distribution is part of what we measure. Sentences are split on terminal punctuation followed by whitespace, with 32 abbreviations and initials protected so that \emph{e.g.} and \emph{et al.} do not end a sentence. A colon does not end a sentence. Appendix~\ref{app-preprocessing} provides every list and pattern in full.

\subsection{Measures}
\label{sec-method-measures}
We describe our measures below. We first analyzed the corpus on the word and sentence level. Then, we investigated changes in registers, and finally, we explored the change in lexical diversity.

\subsubsection{Word Level} We measured the mean word length in characters, mean syllables per word, share of words of seven characters or more, and share of polysyllabic words. The seven-character threshold is the long-word definition used by the Lix readability index and its simplification Rix~\cite{77e9b71c-fe84-3d01-afb5-fc072e945ca7}. We compute the pre-post contrast at every cutoff from five to nine characters, yielding Cliff's delta~\cite{macbeth2011cliff} between 0.54 and 0.58, compared with 0.57 at seven, so the threshold sets the scale of the measure rather than the size of the effect.

\subsubsection{Sentence Level} We measured the mean sentence length, its standard deviation, and its coefficient of variation. The coefficient of variation is included because the claim that machine-assisted prose is rhythmically uniform is a claim about variance rather than about the mean.

% [MA] "From an NLP perspective, I would expect proper definitions for lexical density,
% hedges, boosters" and "how did you choose these?" and "References?"
\subsubsection{Register} For the register, we measured the lexical density, hedges, boosters, self-mentions, modal verbs, nominalization, and passive voice, each per 10,000 tokens. Hedges, boosters, and self-mentions are three of the four components of stance in Hyland's model of academic interaction~\cite{hyland2018metadiscourse}, where stance is the way an author signals how firmly they stand behind a claim and how visibly they are present in the text. Hedges soften a claim, such as \emph{may}, \emph{suggest}, and \emph{likely}. Boosters strengthen one, such as \emph{clearly}, \emph{demonstrate}, and \emph{show}. Self-mentions are first-person references to the authors themselves, such as \emph{we} and \emph{our}. Modal verbs, such as \emph{can}, \emph{may}, \emph{must}, and \emph{should}, mark possibility and obligation. Lexical density is Ure's measure~\cite{ure1971lexical}, the share of tokens that carry lexical rather than grammatical content, computed here against a 161-item closed-class list so that it is the exact complement of the function-word share. We use Ure's denominator of total words rather than Halliday's later revision of clauses. Nominalization turns a process or quality into a noun, as when \emph{analyze} becomes \emph{analysis} or \emph{complex} becomes \emph{complexity}. It belongs to a third line of work, on the phrasal rather than clausal complexity of academic prose~\cite{BIBER20102}, which also supplies the mechanism we return to in the discussion. Hedges, boosters, self-mentions, and modals are counted by membership of fixed word lists reproduced in Appendix~\ref{app-preprocessing}, not by any judgment of what a word is doing in context. We therefore measure the availability of a stance marker rather than its function, and a hedge inside a quotation counts the same as one in the authors' own voice. Nominalization is detected by suffix, and passive voice by a form of \emph{be} followed within two tokens by a past participle, as in \emph{the data were collected} rather than \emph{we collected the data}. We also record further rates for LLM marker words such as em dash, en dash, colon, semicolon, and comma, and compute Flesch Reading Ease~\cite{flesch1948}, Coleman-Liau~\cite{colemanliau1975}, and the Automated Readability Index~\cite{ari1967} for comparability with prior work.

\subsubsection{Lexical Diversity} We measured five aspects of how varied the vocabulary is, that is, how far a text draws on a wide range of words rather than reusing a small set. The type-token ratio~\cite{Richards_1987} is the count of distinct word types divided by the total number of words, so a higher value means a more varied vocabulary. It is the simplest such measure, but it falls mechanically as a text grows longer. Moving-Average Type-Token Ratio (MATTR)~\cite{mattr2010} removes that length dependence by averaging the type-token ratio over a sliding 500-token window. Measure of Textual Lexical Diversity (MTLD)~\cite{mtld2010} captures how far the text runs, on average, before its running vocabulary begins to repeat, computed forward and backward and averaged; a higher value indicates that diversity is sustained across longer stretches. Yule's K~\cite{yule1944} instead measures how heavily a text leans on a few frequently repeated words, where a higher value means less diversity. Shannon entropy~\cite{shannon1948} treats the word-frequency distribution as a probability distribution and measures how evenly the text spreads across its vocabulary, with higher values for a flatter, more varied distribution. Because the type-token ratio falls with length\footnote{Median body length grows from 2,704 tokens in 1982 to 10,973 in 2026.}, and to place all five measures on the same baseline, each is also computed on a fixed opening window of the first 1,500 tokens. That window is not the same part of every paper. It covers 41\% of a typical paper from 1982 to 1995 and 15.5\% of one from 2018 to 2026, so the early years are matched on most of their text and the recent years on their opening sections, where the vocabulary is not representative of the whole. We decided to report both for completeness.

% [MA] "In 4.4 you argue that the break point is 2024. Maybe flip the two subsections."
% Break estimation now precedes the pre-post test, so 2024 is a consequence, not an assumption.
\subsection{Metric Changes}
\label{sec-method-baseline}
We first applied all the analyses to the entire CHI corpus. For each measure, we take the yearly median across documents, yielding one value per year and making it robust to the heavy right tail in lengths and rates. We then fit two independent straight lines to that series, one on each side of a break year, and choose the break that minimizes their combined residual sum of squares. Fitting the two lines independently allows both the slope and the level to change at the break, not just the slope. Each segment must span at least three years, and the calendar year is the predictor, so the missing 1984 edition enters as a gap rather than a compressed step. The break year is estimated from the data rather than fixed at 2024, when LLMs became mainstream as writing assistants for CHI papers. We assess how well it is localized by resampling documents within each year with replacement, recomputing every year's median, and refitting over 400 bootstrap replications. This carries the sampling uncertainty in each year's median through to the break estimate.

\subsection{Post-2024 Period}
\label{sec-method-phase3}
The LLM post-period is 2024 to 2026, which is every edition whose writing could have drawn on widely available generative AI tools, since CHI 2024 was submitted in September 2023. The only open choice is how much history to contrast it against, and we match it with the three years immediately before, which include 2021 to 2023. The selected CHI proceedings keep the two groups comparable in size. Furthermore, almost every measure at CHI has drifted over the past four decades, so a comparison between 1982 and 2023 would mostly reflect that long-term trend rather than any recent change due to LLMs. And the three years just before 2024 already carry the fastest part of that drift, so they are a demanding baseline, and a shift that still stands out against them is not simply the trend continuing.

At these sample sizes, almost any difference reaches significance. Thus, we report Cliff's delta computed from the Mann-Whitney statistic, with Holm-corrected p-values. The full record then reenters as the benchmark. Every prior three-year block in CHI is compared with the three years before it in the same way, yielding 37 prior three-year shifts for each measure. The 2021 - 2023 to 2024 - 2026 shift is then ranked against those 37.

\begin{figure*}[t]
  \centering
  \includegraphics[width=\textwidth]{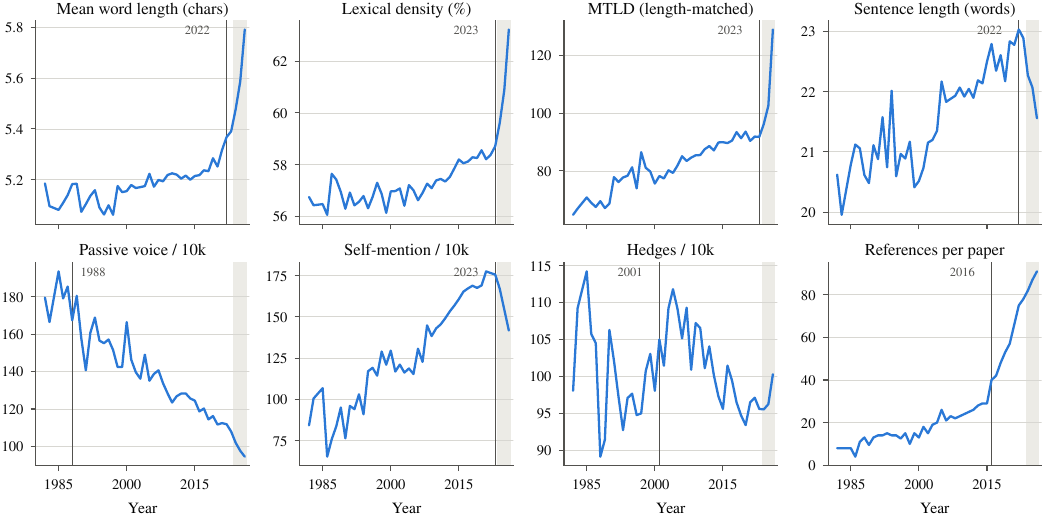}
  \caption{Year-level medians over all 14,262 archival contributions. Each panel plots one measure against the calendar year, from 1982 to 2026, with its own vertical scale, so panels show shape and timing rather than a common unit. The vertical line in each panel marks the estimated break year from the data, and the shaded band indicates 2024 to 2026, when LLMs were available at scale. The eight measures are mean word length, the average length of a word in characters; lexical density, the percentage of words carrying content rather than grammatical function; MTLD, a length-robust measure of vocabulary variety, higher meaning more varied; sentence length in words; passive voice, self-mention, and hedges, each a rate per 10,000 words for be-passives, first-person reference to the authors, and claim-softening words such as \emph{may} and \emph{suggest}, and references per paper. Word-level complexity is flat within 0.2 characters from 1982 to 2020 and breaks in the early 2020s. Passive voice halves across the period, self-mention doubles and then reverses, and hedging moves within a narrow band throughout.}
  \label{fig-baseline}
  \Description{Eight small line charts sharing a year axis from 1982 to 2026. Mean word length, lexical density, and MTLD are nearly flat until about 2020 and then rise steeply. Sentence length rises gradually, then falls after 2022. Passive voice declines steadily from 179 to 95 per ten thousand tokens. Self-mentions rose from 84 to 178 by 2021, then fell to 142. Hedges oscillate between 89 and 114 with no trend. References per paper rise from 8 to 91.}
\end{figure*}

\subsection{Separating Writing From Subject Matter and From Format}
\label{sec-method-controls}
\subsubsection{Function-Word Classifier} Closed-class words carry no subject matter, so if the pre- or post-LLM era of a paper can be predicted from them alone, the change cannot be attributed to CHI publishing different research. We fit logistic regression on the relative frequencies of 157 function words per document and report the five-fold cross-validated area under the curve. Three comparisons bound the result. A model using the 300 most frequent content words shows what topic alone achieves, a placebo split at 2014 shows what an arbitrary division of a quiet period achieves, and shuffled labels give the chance floor. We also report the fitted coefficients, since a classifier is only as trustworthy as the features it leans on. 

% [MA] "Burrow's delta is a descriptive metric not a classifier"
% [MA] "How stable are these profiles? (variance within the subsamples)"
\subsubsection{Burrows's Delta} Burrows's Delta compares whole function-word profiles and gives a descriptive distance between two years, not a classification. Year profiles are z-scored across the series and compared by mean absolute difference. Because yearly counts range from 29 to 1,703 CHI papers per year, and a mean profile built from a few documents is noisier and therefore further from the truth, we recompute each profile from a random subsample of 29 documents and average over 60 draws. Only size-matched values are used, and we report the standard deviation across draws to make the stability of each comparison visible. We compute distances both between neighboring years and across 10-year spans. The two measures answer different questions. Language changes gradually, so CHI can travel a long way without any single year-on-year step appearing large, and a design that inspects only adjacent years cannot distinguish slow directional drift from stability.

\subsubsection{Formulaicity} Lexical diversity asks how many distinct words a document uses. Homogenization is a claim about templated phrasing shared across documents, so we measure four-gram reuse across papers rather than repetition within a single document. This measure must be matched on two things at once. Cross-document sharing rises mechanically with the number of documents compared, so a year represented by many short papers would look more formulaic than the same prose split across fewer long ones. We therefore draw a fixed 25 documents per year, truncate each to its first 2,000 tokens, and average over 60 random draws. Those counts are set by the smallest year in the corpus, which is why the series can span all 44 years.

% [MA] "which ones" - naming the affected measure rather than saying "several"
\subsubsection{Historical Markers} Terminology and reporting conventions with known histories serve both as substance and as a check that the pipeline tracks real change. We count occurrences per 10,000 tokens of participant and subject terminology, such as man-machine against human-computer, markers of statistical reporting practice, qualitative method vocabulary, positionality statements, and references to language models.

\subsection{Replicating Published AI-Marker Analyses}
\label{sec-method-markers}

Two published marker sets are applied to the CHI corpus so that our findings can be compared against a cross-disciplinary baseline rather than merely described.

\subsubsection{Lexical Items} The 17 items of Comas-Forgas et al.~\cite{comasforgas2026}, identified by snowball sampling of Scopus abstracts and validated by a growth threshold. We carry their reported Scopus growth ratios so that CHI can be placed against them. Following their procedure, we match on the stem of the word, so \emph{delv} captures delve, delves, delved, and delving.

\subsubsection{Title Verbs} The 15 verbs of Comas-Forgas et al.~\cite{comasforgas2025}, matched in CHI paper titles. Their growth ratio is the 2024 count divided by the mean of the 2021 and 2022 counts. CHI 2024 was submitted in September 2023 and is likewise the first cohort that could have used these tools, so the definition transfers without adjustment. Because our corpus runs two years beyond theirs, we also report the equivalent 2026 ratio. We apply the markers to both body text and abstracts. Prior work measures abstracts, but CHI abstracts are roughly 150,000 tokens per year across CHI, so a marker occurring at 0.05 per 10,000 tokens falls below one occurrence per year, and its growth ratio is undefined. The body text contains 115.5 million tokens and is where these measures can resolve a trend. Control terms are ours rather than theirs, since the published control list could not be recovered from the article. We use five frequent connectives with no claimed association to machine-assisted prose, which bind what the corpus does on its own.

\subsubsection{Separating Marker Use From Marker Study} A paper about language models may quote model output, which would raise marker counts without any change in how the authors write. We therefore split each year into papers whose abstracts or introductions mention language models and those that do not, and compute the marker rates separately.

\subsubsection{Case Analysis: Dash Decomposition} The em and en dash rate is the most quotable signature in this literature. We decompose dash use by context rather than reporting a single rate. We count em dashes and en dashes occurring between two letters, hyphens occurring inside words, and dashes in numeric ranges. A change in writing should move the sentence-level dashes without moving the marks that join compounds or ranges.

\begin{table*}[t]
  \caption{Year-level medians, CHI archival contributions, body text. Rates are per 10,000 tokens. The final column marks the direction over the full period. A single arrow is a monotonic change, a paired arrow rises then falls (\tupdown) or falls then rises (\tdownup), and \tflat\ marks no trend. Color reinforces the arrow direction and is not needed to read it.}
  \label{tab-baseline}
  \small
  \begin{tabular}{@{}lrrrrrrc@{}}
    \toprule
    Measure & 1982 & 1995 & 2010 & 2020 & 2023 & 2026 & Trend \\
    \midrule
    Mean word length      & 5.2 & 5.1 & 5.2 & 5.3 & 5.4 & 5.8 & \tup \\
    Lexical density       & 56.7 & 56.3 & 57.4 & 58.6 & 58.7 & 63.2 & \tup \\
    MTLD, length-matched  & 64.8 & 81.3 & 85.5 & 93.6 & 91.8 & 128.9 & \tup \\
    Sentence length       & 20.6 & 20.6 & 21.9 & 22.8 & 22.9 & 21.6 & \tupdown \\
    Passive voice         & 179 & 155 & 123 & 112 & 108 & 95 & \tdown \\
    Self-mention          & 84 & 117 & 143 & 169 & 176 & 142 & \tupdown \\
    Hedges                & 98 & 98 & 107 & 93 & 96 & 100 & \tflat \\
    Nominalization        & 550 & 503 & 483 & 513 & 552 & 663 & \tdownup \\
    Body tokens           & 2704 & 4946 & 6480 & 7899 & 10669 & 10973 & \tup \\
    References            & 8 & 14 & 24 & 57 & 78 & 91 & \tup \\
    \bottomrule
  \end{tabular}
\end{table*}

\section{Results}
\label{sec-results}
We present our results below. First, we begin with an analysis of registers, followed by the analysis of lexical diversity and AI-typical marker words.

\subsection{44 Years of Slow Drift in the Major Metrics}
\label{sec-results-baseline}

\begin{figure*}[t]
  \centering
  \begin{subfigure}[t]{0.48\linewidth}
    \vspace{0pt}
    \centering
    \includegraphics[width=\linewidth]{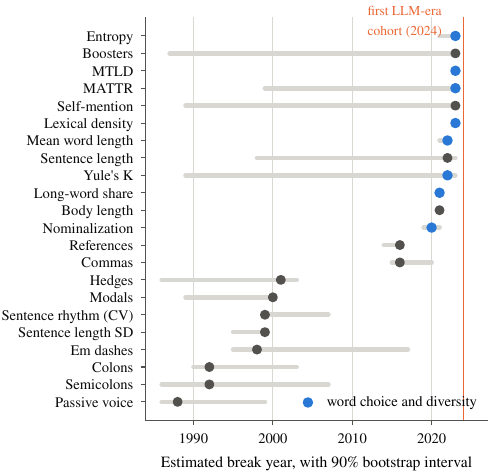}
    \subcaption{}
    \label{fig-changepoints}
    \Description{A dot plot of twenty measures ordered by estimated break year, each with a bootstrap interval. Lexical density, MATTR, MTLD, long-word share, Yule's K, entropy, mean word length, and nominalization cluster between 2020 and 2023, all at or before the 2024 line. Sentence rhythm breaks in 1999, colons and semicolons in 1992, and passive voice in 1988.}
  \end{subfigure}\hfill
  \begin{subfigure}[t]{0.48\linewidth}
    \vspace{0pt}
    \centering
    \includegraphics[width=\linewidth]{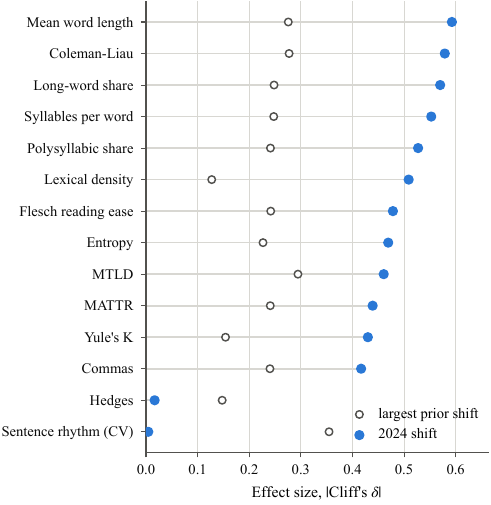}
    \subcaption{}
    \label{fig-effects}
    \Description{A dot plot of fourteen measures ranked by effect size. For twelve measures, including mean word length, Coleman-Liau, long-word share, and MTLD, the recent shift is larger than any previous three-year shift. For hedges and sentence rhythm, the recent shift is much smaller than the largest prior shift.}
  \end{subfigure}
  \caption{\textbf{Left:} When each measure changed. Points show the estimated break year from the data, and bars show the 90\% bootstrap interval; a wide bar indicates that the timing is not well identified. Word choice and lexical diversity cluster in 2021 to 2023, at or before the first cohort that could have used generative tools. Sentence-level and punctuation measures break down on their own, far earlier in their timelines. \textbf{Right:} The 2021 to 2023 against 2024 to 2026 shift, benchmarked against every prior three-year shift in the history of CHI. Filled marks give the recent shift, and hollow marks the largest shift previously recorded for that measure. The twelve largest all exceed anything in the past of CHI. The two at the foot do not.}
  \label{fig6-fig2}
\end{figure*}

\autoref{fig-baseline} and \autoref{tab-baseline} give the year-level medians over all 14,262 archival contributions. Our results show that CHI changed slowly and steadily for four decades across all registers. Passive voice fell by almost half, from 179 occurrences per 10,000 tokens in 1982 to 95 in 2026, declining since the late 1980s and owing nothing to any recent event. Self-mentions doubled between 1982 and 2023, from 84 to 176, and then reversed. Nominalization is U-shaped, falling to 2015 and rising since. Reference lists grow elevenfold, from a median of 8 to 91, and body length grows fourfold. The 2010 column ranks below its neighbors in length and citation count because the Notes track is included here and spans 2006 to 2017.  Mean word length stayed within 0.2 characters around 5.2 for the entire period from 1982 to 2020. The estimated break is 2021 for long-word share, with bootstrap support of 0.95, and 2022 for mean word length. Sentence-level and punctuation measures break down on their own, much earlier than timelines. Sentence-length variability changes in 1999, with a  0.95, and the punctuation measures in the early 1990s. \autoref{fig-changepoints} gives every estimated break year with its bootstrap interval, and the separation between the two clusters is the clearest single statement of the timing.

The onset of the word-level rise precedes the first cohort that could have used generative tools by at least one conference cycle, which is a clean null. It moves within a narrow band for 44 years, and its estimated break carries bootstrap support of 0.24 over an interval spanning 1986 to 2003, which is to say none. This is notable given how central hedging is to accounts of scholarly voice.

\subsection{Metric Changes Before and After LLMs}
\label{sec-results-phase3}

\begin{figure*}
  \centering
  \begin{minipage}[t]{0.56\linewidth}
     \vspace{0pt}
      \centering
      \includegraphics[width=\columnwidth]{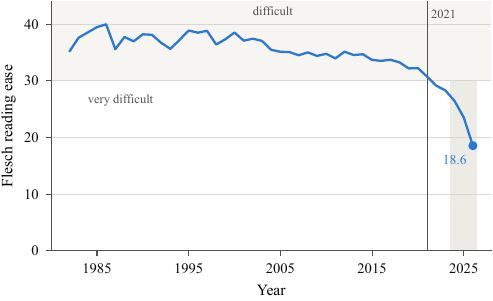}
      \caption{Flesch reading ease by year, with the difficulty bands of the scale drawn behind the series and the estimated break year marked. CHI prose is inside the band the scale
      calls difficult for its first four decades and drops below it in 2022. The fall from 23.6 in 2025 to 18.6 in 2026 exceeds the entire drift from 1982 to 2020. Reading ease is reported
      here as a summary of word and sentence length.}
      \Description{A line chart of Flesch reading ease from 1982 to 2026. The series oscillates between 35 and 40 until about 2005, drifts down to 32 by 2020, and then falls steeply through a marked break at 2021 to 18.6 in 2026. A horizontal band boundary at 30 divides the region labeled difficult from the region labeled very difficult, and the series crosses it in 2022.}
      \label{fig-flesch}
  \end{minipage}\hfill
  \begin{minipage}[t]{0.40\linewidth}
    \vspace{0pt}
    \centering
    \captionsetup{type=table}
    \caption{Pre-period 2021-2023 against post-period 2024-2026, CHI archival contributions. Delta is Cliff's delta, positive meaning an increase. The final column ranks the shift against all 37 prior three-year shifts in the history of CHI.}
  \label{tab-prepost}
  \small
  \begin{tabular}{@{}lrrrl@{}}
    \toprule
    Measure & Pre & Post & Delta & Rank \\
    \midrule
    Mean word length     & 5.36 & 5.63 & $+$0.59 & largest \\
    Coleman-Liau         & 14.45 & 16.01 & $+$0.58 & largest \\
    Long-word share      & 33.1 & 37.0 & $+$0.57 & largest \\
    Lexical density      & 58.4 & 61.4 & $+$0.51 & largest \\
    Flesch reading ease  & 29.4 & 22.3 & $-$0.48 & largest \\
    MTLD, matched        & 91.4 & 108.6 & $+$0.46 & largest \\
    Yule's K, matched    & 95.9 & 83.7 & $-$0.43 & largest \\
    Nominalization       & 541 & 628 & $+$0.33 & largest \\
    Boosters             & 42.5 & 49.4 & $+$0.23 & largest \\
    Self-mention         & 176.6 & 152.8 & $-$0.22 & largest \\
    Passive voice        & 110.8 & 97.1 & $-$0.21 & largest \\
    Sentence length      & 22.89 & 21.88 & $-$0.19 & largest \\
    Hedges               & 95.9 & 97.8 & $+$0.02 & 24th pct \\
    Sentence-length CV   & 0.54 & 0.54 & $-$0.004 & 3rd pct \\
    \bottomrule
  \end{tabular}
  \end{minipage}
\end{figure*}

Comparing 2021-2023 (n = 2,262) with 2024-2026 (n = 4,009), every word-level and diversity measure moves substantially. Mean word length rises 5.0\% with a Cliff's delta of 0.59, long-word share rises 11.8\% with a delta of 0.57, lexical density rises 5.1\% with a delta of 0.51, and Flesch reading ease falls 24.1\%. Ranked against all 37 prior three-year shifts in the history of CHI, our measures show high shifts (see \autoref{fig-effects} and \autoref{tab-prepost}).

Reading ease is worth stating on its own scale, since it is the only measure here with a published interpretation rather than a purely relative one (see \autoref{fig-flesch}). It holds between 35 and 40 for the first quarter century, drifts down to 32.3 by 2020, then falls to 30.8, 29.2, 28.3, 26.5, 23.6, and 18.6 over the six years to 2026. CHI spent its first four decades inside the band, the scale calls it difficult, and dropped below it in 2022, and the single-year fall from 2025 to 2026, of 5.0 points, is larger than the entire drift of 2.9 points from 1982 to 2020. Coleman-Liau and the Automated Readability Index move the same way over the same years, so this is not an artifact of the syllable counter that Flesch depends on
and the other two do not.

Two results run against the homogenization account. Vocabulary becomes more varied rather than less, with length-matched MTLD rising 18.9\% and Yule's K falling 12.7\%. And sentence rhythm does not become more uniform. The hedging coefficient likewise does not move, with a delta of 0.02 at p = 0.81 after correction. The recent shift ranks in the third percentile of the history of CHI, meaning that almost every prior three-year period moved more.

% [MA] both robustness objections answered with data rather than argument
\subsubsection{Paper Length as Confound} CHI has moved to a word limit since CHI 2021, abolishing page limits. The change to a word limit is reflected in an increase in the number of tokens. CHI papers grew from a median of 7,899 body tokens in 2020 to 10,973 in 2026. But within the 2021 to 2026 documents, body length correlates with mean word length at 0.05 and with long-word share at 0.03. Splitting those documents into quartiles of body length and recomputing the contrast inside each, mean word length rises by 0.273, 0.290, 0.253, and 0.262 characters from the shortest quartile to the longest, against 0.268 pooled. The shift is the same size in papers with 3,286 tokens and in papers with 27,018 tokens. Computed on full body text instead of the first 1,500 tokens, MTLD rises 17.7\% against 18.9\%, MATTR rises 6.1\% against 5.3\%, Yule's K falls 14.4\% against 12.7\%, entropy rises 2.8\% against 1.9\%, and the type-token ratio rises 7.8\% against 6.6\%. Every measure agrees in sign and in rough magnitude. However, we noticed one limitation. The change to a word limit falls in 2021, the same window as the estimated word-level break, so the format change and the onset coincide in time. The quartiles show that length is not the mechanism behind the shift, since the effect holds at every body length, but we cannot separate a discrete format effect from any tool effect at the break itself.

\subsection{Classifying Papers into the Pre- and Post-LLM Era}
\label{sec-results-classifier}

Whether a paper was published before or after 2024 can be predicted from its function words alone, with a cross-validated area under the curve of 0.864. Function words carry no meaning, so this cannot be explained by CHI publishing more research about artificial intelligence. The classifier is L2-regularized logistic regression over per-document word rates, standardized so that coefficients are comparable across words of very different frequencies, and each AUC is the mean over five-fold cross-validation, with the standard deviation reported across folds. Each split uses a symmetric six-year window, three years on each side, so the real split trains on 2021 to 2026, divided at 2024.

The comparison models bound the claim. A model built from the 300 most frequent content words achieves 0.857, so function words carry as much of the era signal as the topic does. A placebo split in 2014 reached only 0.581. The placebo is the same procedure applied to a boundary where nothing is claimed to have happened: its own six-year window, 2011 to 2016, divided at 2014. If the classifier separated those years as sharply as it separates 2024, the method would be manufacturing a difference wherever it cuts, and the 2024 result would be an artifact of the procedure rather than a real change. It does not, so the strong 2024 split is credible. The placebo draws on fewer documents, 2{,}664 against 6{,}271, only because CHI was smaller a decade earlier, and it still lands near chance despite that. Shuffled labels give 0.498, exactly at chance. \autoref{fig-classifier} shows all four.

A classifier is only as trustworthy as the features it leans on, so we report them. Nothing at the top of the list looks like leaked subject matter. The words most raised in 2024 to 2026 are \emph{across}, \emph{this}, \emph{into}, \emph{where}, \emph{while}, and \emph{beyond}, and the words most lowered are \emph{of}, \emph{a}, \emph{is}, \emph{be}, \emph{to}, and \emph{were}.

Two of those patterns are interpretable, and one corroborates a result reported elsewhere. Also among the lowered words are \emph{were} and \emph{had}, the past tense receding, and \emph{we}, which matches the decline in self-mention in \autoref{tab-prepost}. More telling is the sharpest single coefficient, the fall of \emph{of}, accompanied by \emph{a}, \emph{an}, \emph{is}, and \emph{are}. That is the signature of phrasal compression, in which modification migrates inside the noun phrase and \emph{adaptation of the user interface} becomes \emph{user interface adaptation}. Compression predicts exactly the decomposition we observe elsewhere, longer words inside shorter sentences, and it is reached here by a route that shares no measure with the one that produced it.

\begin{figure*}[t]
  \centering
  \begin{subfigure}[t]{0.48\linewidth}
    \vspace{0pt}
    \centering
    \includegraphics[width=\linewidth]{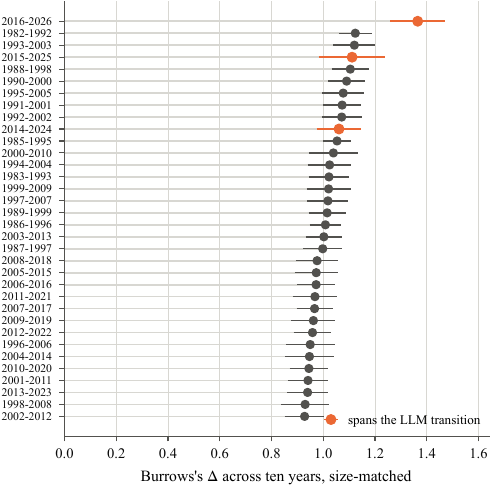}
    \subcaption{}
    \label{fig-burrows-long}
    \Description{A dot plot of twenty measures ordered by estimated break year, each with a bootstrap interval. Lexical density, MATTR, MTLD, long-word share, Yule's K, entropy, mean word length, and nominalization cluster between 2020 and 2023, all at or before the 2024 line. Sentence rhythm breaks in 1999, colons and semicolons in 1992, and passive voice in 1988.}
  \end{subfigure}\hfill
  \begin{subfigure}[t]{0.48\linewidth}
    \vspace{0pt}
    \centering
    \includegraphics[width=\linewidth]{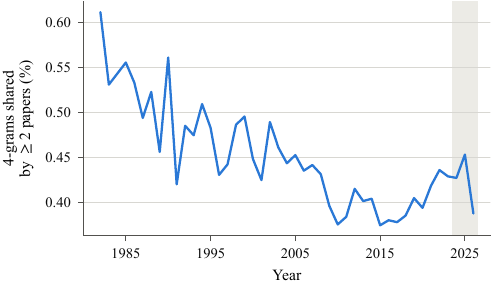}
    \subcaption{}
    \label{fig-formulaicity}
    \Description{A dot plot of fourteen measures ranked by effect size. For twelve measures, including mean word length, Coleman-Liau, long-word share, and MTLD, the recent shift is larger than any previous three-year shift. For hedges and sentence rhythm, the recent shift is much smaller than the largest prior shift.}
  \end{subfigure}
  \caption{\textbf{Left:} Burrows's Delta across ten-year spans, every decade in the history of CHI ranked together, with bars giving the standard deviation over 60 size-matched draws. The 2016-to-2026 span is the largest on record, and its interval does not reach the runner-up. \textbf{Right:} Shared phrasing across papers, measured as the share of four-gram types occurring in at least two distinct papers. Matched on both document count and length, 25 papers per year were truncated to 2,000 tokens each and averaged over 60 draws, because cross-document sharing otherwise rises with the number of documents compared. Templated phrasing declines from the 1980s to about 2009, holds between 0.37\% and 0.41\% through the following decade, and then rises. Every year from 2021 to 2025 exceeds that decade's maximum, before 2026 returns to it. An earlier version of this series showed a decline across the last three editions, which turned out to be publisher boilerplate rather than prose.}
  \label{fig14-fig4}
\end{figure*}

\subsection{Investigating Years with Lexical Style Changes}
\label{sec-results-delta}
\begin{figure*}[t]
  \centering
  \begin{subfigure}[t]{0.49\textwidth}
    \centering
    \includegraphics[width=\linewidth]{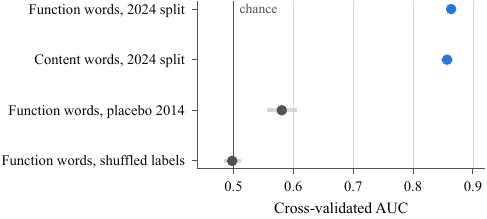}
    \Description{A dot plot of four models. Function words at the 2024 split reach 0.864, and content words 0.857. A placebo split at 2014 reaches 0.581 and shuffled labels 0.498, at chance.}
    \subcaption{}
    \label{fig-classifier}
  \end{subfigure}\hfill
  \begin{subfigure}[t]{0.49\textwidth}
    \centering
    \includegraphics[width=\linewidth]{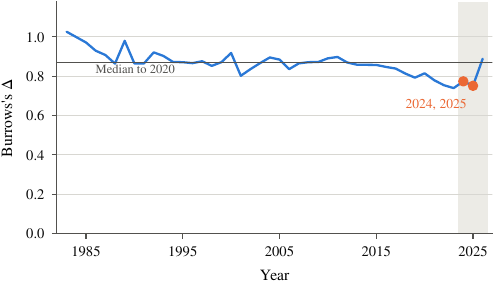}
    \Description{A line chart of Burrows's Delta between consecutive years from 1983 to 2026. Values fell from about 1.0 in the mid-1980s to about 0.80 by 2015 and dropped further after 2020. The 2024 and 2025 points, highlighted, sit below the historical median line of 0.864. Only 2026 rises again, to 0.883.}
    \subcaption{}
    \label{fig-burrows}
  \end{subfigure}
  \caption{\textbf{Left:} Whether a paper appeared before or after 2024, predicted from its vocabulary. Points give five-fold cross-validated AUC, and bars give the standard deviation across folds. Function words carry as much of the signal as content words do, and function words hold no subject matter. A placebo split in 2014 and shuffled labels bound the result from below. \textbf{Right:} Burrows's Delta between consecutive years, computed on whole function-word profiles rather than on measures chosen in advance. Every year is subsampled to the size of the smallest year, 29 documents, and averaged over 60 draws, so that sparse early years are not made to look turbulent by sample size alone. Every transition from 2021 onward except 2026 ranks in the bottom five of 43. Read alone, this suggests stability.}
  \label{fig-classifier-burrows}
\end{figure*}

Between consecutive years, the recent period looks placid. With every year subsampled to a common 29 documents, the distance between neighboring years has a median of 0.87 up to 2020 and 0.76 from 2021 onward, and every recent transition except one ranks in the bottom five of 43 (see \autoref{fig-burrows}). The 2022-to-2023 step is the smallest in the history of CHI. We find that language drifts gradually, so CHI can travel a long way without any single year-on-year step looking large. Measured across ten years rather than one, and size-matched the same way, the 2016-to-2026 span is the largest stylistic shift in the history of CHI, at a delta of 1.365 and 2.3 standard deviations clear of the runner-up (see \autoref{fig-burrows-long}). It exceeds 1982-to-1992 at 1.124 and 1993-to-2003 at 1.120, and the three spans that reach the LLM transition occupy ranks one, four, and ten of 34. The change is slow, so the year-on-year steps stay small, and it runs across many dimensions at once, so it accumulates. That is exactly the profile a year-on-year design cannot capture, and it is why a classifier separates recent papers from older ones at 0.864, with no single year looking unusual.

\subsection{The Rise of Templated Phrasing as Indicator for Homogenization}
\label{sec-results-formulaicity}
The share of four-gram types occurring in at least two papers falls unevenly from 0.61\% in 1982 to a mean of 0.39\% across 2009 to 2020, holding between 0.37\% and 0.41\% through that decade (see \autoref{fig-formulaicity}). It then rises. Every year from 2021 to 2025 is above that decade's maximum, at 0.42\%, 0.44\%, 0.43\%, 0.43\%, and 0.45\%, before 2026 returns to 0.39\%, inside the old band. This is a modest rise on the measure most directly aimed at the homogenization claim, and we report it as such. The increase is about a tenth of the preceding decade's level, it holds across five consecutive editions, and it is not shared subject matter, since removing every four-gram that contains artificial-intelligence vocabulary leaves it intact at 0.049 percentage points against 0.046. Two things still argue against reading it as generative tools homogenizing prose. The rise begins in 2021, three editions before the first cohort that could have used those tools. And 2026, the edition where such use should be most widespread, is the one year that returns to the historical band. Both corrections behind this series moved it, and we record them because each would affect any replication. The first was a matching failure. Holding only a token budget constant, let the document count fall from 31 in early years to 9 in recent ones, and since cross-document sharing rises with the number of documents compared, this manufactured a decline. The second was contamination. Even after matching, the most-shared four-grams of CHI 2020 were phrases from the ACM permission notice, present in 20 of the 25 papers drawn, while CHI 2019 carried the notice in only 6\% of its papers. The measure was reading ACM production practice rather than authorship.

\subsection{AI-Typical Marker Words}
\label{sec-results-markers}

\begin{figure*}[t]
  \centering
  \includegraphics[width=\textwidth]{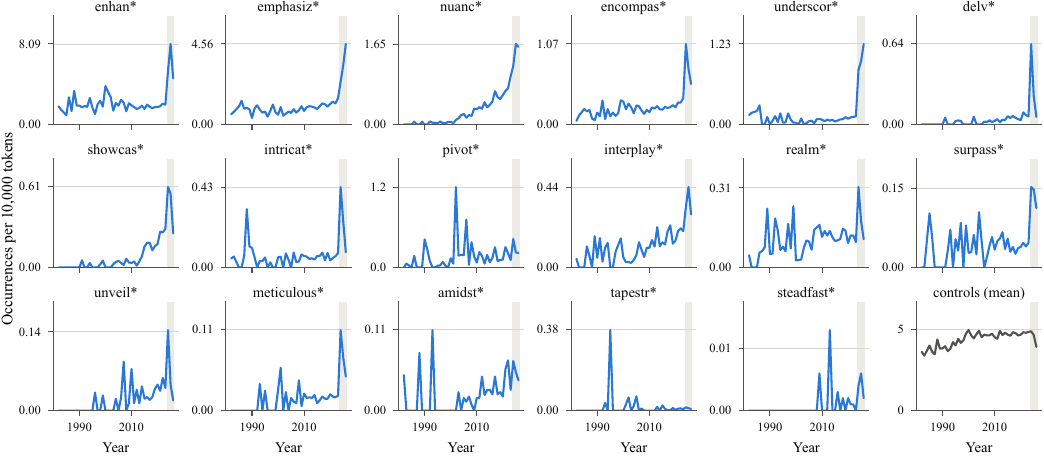}
  \caption{Each of the 17 AI-associated lexical items~\cite{comasforgas2026} across the whole corpus, on its own vertical scale, because their raw frequencies differ by three orders of magnitude. The final panel gives the mean of the five control terms. The shaded band marks the 2024 to 2026 cohorts. Two things are visible here that a grouped plot hides. Several items rise and then fall back within the band rather than climbing, and several were never part of this CHI vocabulary in the first place, with \emph{steadfast} reaching 0.013 occurrences per 10,000 tokens at its highest and
  \emph{tapestry} appearing only as the name of a system studied in 1995.}
  \Description{A grid of eighteen small line charts from 1982 to 2026, one per marker plus one for the control average. Enhance, emphasize, nuance, encompass, underscore and delve show clear rises in the shaded 2024 to 2026 band. Delve, intricate, showcase, realm, and unveil rise then fall within that band. Steadfast stays near zero throughout and tapestry has a single spike in 1995 and little since. The control average ranges from 3.4 to 5.0 with no trend over the entire period.}
  \label{fig-marker-grid}
\end{figure*}

Applying the 17 lexical items of Comas-Forgas et al.~\cite{comasforgas2026} to CHI body text reproduces their central result and then extends past it. Eleven of the 17 markers more than double between the 2021 and 2022 baseline and 2024, while all five control terms stay flat or decline, between 0.76 and 1.03. The rise is real, and it is not a corpus-wide inflation. \autoref{fig-marker-grid} shows each marker separately across the whole corpus, which also makes clear that the published set does not transfer wholesale to a single venue. \emph{Steadfast} reaches 0.013 occurrences per 10,000 tokens at its highest. \emph{Tapestry} has a single spike of 0.38 in 1995, when a collaborative filtering system of that name was a subject of study rather than a figure of speech, and otherwise is below 0.07 for the whole corpus. Two of the seventeen are therefore effectively absent from HCI writing and can carry no signal here.

\begin{figure*}
  \centering
  \includegraphics[width=\textwidth]{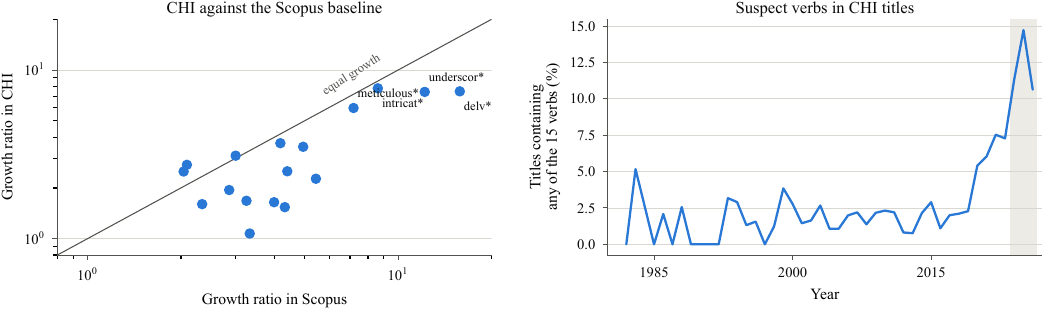}
  \caption{\textbf{Left:} CHI growth ratios against the Scopus values reported by Comas-Forgas et al.~\cite{comasforgas2026}, both on logarithmic axes, with the line of equal growth. Almost every marker falls below it. \textbf{Right:} the share of CHI titles containing at least one of the 15 suspect verbs of Comas-Forgas et al.~\cite{comasforgas2025}.}
  \Description{Two panels. The left panel is a log-log scatter of 17 markers showing the CHI growth ratio against the Scopus growth ratio, with most points below the diagonal. The right is a line chart from 1982 to 2026 of the percentage of titles containing a suspect verb, below 4 percent until about 2019, then rising steeply to 14.7 percent in 2025 and falling to 10.6 percent in 2026.}
  \label{fig-replication}
\end{figure*}

One group spikes in 2024 and returns to its pre-LLM level within two years. The word \emph{delve} runs at 0.07 occurrences per 10,000 tokens in 2022, rises to 0.64 in 2024, and falls to 0.06 in 2026. \emph{Intricate}, \emph{realm}, \emph{unveil}, and \emph{showcase} follow the same path. A second group rises and stays. \emph{Underscore} moves from 0.12 to 0.83 and then to 1.23, and \emph{emphasize}, \emph{nuance}, and \emph{enhance} behave likewise. Counted at 2026 rather than 2024, only 6 of the 17 markers still show a doubling. A marker set validated on 2024 data, therefore, has a short shelf life, which matters for anyone proposing to use such lists to detect machine-assisted writing. Splitting each year into papers that mention language models and those that do not, the 2024 rise appears in both at almost the same magnitude. In papers that never mention them, \emph{delve} runs at 0.59 per 10,000 tokens in 2024 against 0.76 in papers that do, \emph{showcase} at 0.63 against 0.54, and \emph{underscore} at 0.82 against 0.86. The change is therefore in how CHI writes, not in what a subset of it quotes. Relative to the Scopus baseline, the growth ratios of CHI are systematically below the published values (see \autoref{fig-replication}). The same markers rise in CHI, but less sharply than across Scopus as a whole, so the cross-disciplinary ranking does not transfer cleanly to a single CHI proceeding.

The 15 title verbs by Comas-Forgas et al.~\cite{comasforgas2025} show the same profile. The share of CHI titles containing at least one of them stays below 4\% for the first thirty-seven years of the corpus, with a median of 1.6\% between 1990 and 2018. It then reaches 5.4\% in 2020 and 6.0\% in 2021, rises to 11.4\% in 2024 and 14.7\% in 2025, and falls back to 10.6\% in 2026. The rise, therefore, begins around 2020, four years before the first cohort that could have used generative tools. 

\subsection{Case Analysis: En \& Em Dashes}
\label{sec-results-dashes}

The em dash is the most quotable signature in this literature and the one most prone to measurement error due to its different encodings. Thus, we are reporting it here separately. The em dash rate is exactly zero from 1982 to 1990 and again from 1996, 1998, and 2000, while from 1991 to 1995 and 1997 it runs between 0.24 and 0.45 with no relation to its neighbors. The double hyphen behaves inversely and just as erratically, between 0.07 and 0.89. Neither series tracks writing. Both track which volume was scanned and how it was scanned.

\begin{figure*}[t]
  \centering
  \includegraphics[width=\textwidth]{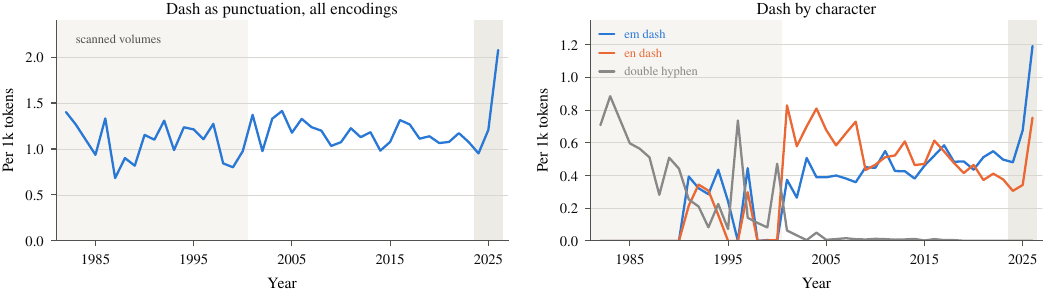}
  \caption{Dash use from 1982. \textbf{Left:} A glyph-independent family counting em dash, en dash, double hyphen, and spaced hyphen between words, which is comparable across the whole period. \textbf{Right:} The same data split by character, with the scanned era shaded. Optical character recognition sometimes renders an em dash as the character and usually as a double or spaced hyphen, so the character count drops to zero in years where nothing about the writing changed and cannot be read across 2001. The transition itself demonstrates the point, since the family measure barely moves while the double hyphen collapses and the em and en dashes stabilize.}
  \Description{Two line charts from 1982 to 2026. The left shows dash punctuation across all encodings, holding between about 0.68 and 1.41 per thousand tokens for forty-four years, then rising to 2.08 in 2026. The right splits the same data by character. Before 2001, the em and en dash series dropped repeatedly to zero while the double hyphen runs high. After 2001 the double hyphen collapses to near zero, the em and en dash series settle between 0.3 and 0.8, and both rise sharply in 2026.}
  \label{fig-dashes}
\end{figure*}

Measured in a way that does not depend on the glyph, by counting em dash, en dash, double hyphen, and spaced hyphen between words as one encoding family, dash punctuation is remarkably stable. It ranges from 0.68 to 1.41 occurrences per 1,000 tokens over the forty-four years to 2025 (see \autoref{fig-dashes}). The 2001 transition is the control case for this measure. The family barely moves across it, while the double hyphen collapses from 0.47 to 0.06 and the en dash rises from 0.005 to 0.83. That is what a pure change of encoding looks like. The 2026 rise does not look like that. The family measure moves from 1.21 in 2025 to 2.08 in 2026, above every previous year in the corpus. A change in encoding redistributes dashes between glyphs and therefore cannot move a measure that already counts every glyph, so the rise cannot be explained as the 2001 transition can. Reading the em dash character alone would suggest a typographic artifact. Reading the encoding-independent family shows a real increase, and that is the reading we adopt.

Two qualifications remain. The rise rests on a single edition. And intra-word hyphens rise by 53\% between 2020 and 2026, from 12.3 to 18.8 per 1,000 tokens, which is consistent with compound modifiers such as human-AI and LLM-based and is therefore a matter of subject matter rather than style. Dash punctuation and compound hyphenation both increased, and only the first is a claim about writing.

\subsection{Limitations}
\label{sec-results-caveats}
Several measures move most in 2026 alone, and two facts point to production rather than prose. Em dash and en dash rates both roughly double in 2026, and a change in writing should lift em dashes without lifting en dashes, which mark numeric ranges. We addressed this specific concern, in which an encoding-independent measure shows that the dash rise survives it. The general concern does not go away, since 2026 rests on a single edition, so we report no 2026-specific magnitude as a headline finding and treat that edition as provisional until a later one confirms it.

The period contains at least four concurrent changes. CHI nearly tripled in size, the pool of people publishing in it turned over, its subject matter shifted decisively toward artificial intelligence, and generative writing tools became available. Our controls rule out some readings but not others. The function-word result indicates that the change is not merely a shift in topic, and the marker analysis shows that the 2024 spike is not confined to papers on language models. Neither separates the arrival of generative tools from the growth of CHI, nor does the onset of the word-level rise in 2021 precede the tools by more than one conference cycle.

Authorship is the confound we control least. A venue that triples in size draws in new authors, and new authors write differently from established authors for reasons unrelated to any tool. A within-author design would hold that constant, and we do not run one. What limits the damage is the length of the series rather than a control. The author pool of CHI has turned over repeatedly across 44 editions, and no prior turnover produced shifts of the size recorded in 2024 to 2026, as established by the benchmark against every prior three-year shift. That makes turnover an unlikely sole explanation without ruling it out as a contributing one. 

Our change-point model fits a single break. Over 44 years, there are plausibly several, and our own estimates fall in two visible clusters rather than one. Second, the pre-period spans three editions, and the entire prior series is 41, so a change that began earlier, with automated language editing rather than generative models, would still register as a post-2024 difference. Benchmarking against every prior three-year shift limits this, but does not remove it. Third, in the marker analysis, we separate papers that mention language models from those that do not; a cleaner design would detect quoted model output and remove it, since an author who quotes model output is plausibly also more willing to write with one. Fourth, our measure of shared phrasing counts a four-gram once it appears in two papers, which treats a phrase shared by two papers and a phrase shared by two hundred as the same event. A weighted version would say more about whether convergence is local or community-wide, and we consider it the most valuable of the four to fix.

\section{Discussion}
We analyzed how writing has changed at CHI, focusing on lexical changes after the public availability of LLMs. We discuss the implications of our results below. We begin by reflecting on the research questions. Then, we discuss the implications of LLM use on the writing of the CHI community.

\subsection{The Lexical Properties of CHI Writing Changed}
\textbf{RQ1} asked how lexical properties changed throughout CHI with an emphasis on the pre- and post-LLM period. On most measures, the change was already slow and long-running before the public availability of LLMs. Papers grew from a median of 2,704 words in the 1980s to 10,973 in 2026, likely due to the extension and removal of page limits. Passive voice almost halved, from 179 per 10,000 tokens to 95, self-mention nearly doubled from 84 to 176, and reference lists grew. Reading ease was 37.7 through the 1980s and 1990s, slipped to 36.0 and then 34.0, and reached an average value of 27.0 in the years 2020 to 2026. However, word choice stands out for how long it stayed still. Mean word length sat within 0.2 characters of 5.2 from 1982 to 2020, then broke in the early 2020s. Register drifted for four decades, and word choice and vocabulary changed sharply and only recently.

This register trend is not specific to CHI. Corpus studies show that academic prose has grown denser and more compressed over decades, packing information into longer noun phrases rather than longer clauses~\cite{BIBER20102, biber1988}, and the direction we see runs against the idea that scholarly writing is becoming more informal~\cite{hylandjiang2017}. The rise in self-mention aligns with the literature, which treats first-person reference and stance as a normal and growing part of academic voice rather than a lapse~\cite{hyland2005stance, hyland2018metadiscourse}. What our corpus adds is measured the same way for 44 years, which shows that the recent word-level break is atop this older drift rather than causing it.

\subsection{Comparing CHI 2021-2023 and 2024-2026 to the History of CHI}
\textbf{RQ2} asked how writing changed at CHI. We analyzed three proceedings before and after the public release of ChatGPT, and compared the shifts against the whole corpus. In other words, we compared the three pre-LLM editions (i.e., 2021 to 2023) against the three post-LLM editions (i.e., 2024 to 2026) and ranked that shift against every prior three-year-to-three-year shift at CHI, resulting in a comparison across 37 proceedings. In this context, our measures record their largest shift ever. Widening the window to a decade instead of a three-year shift tells the same story on whole function-word profiles, where the 2016-to-2026 span is the largest in the history of CHI, standing 2.3 standard deviations clear of the runner-up. Yet year to year the recent steps are among the smallest ever, with 2022-to-2023 the smallest of all.

The interval is why prior estimates disagree with each other and with us. Studies that compare a year or two around the release of ChatGPT report a sharp recent jump in LLM-associated language~\cite{liang2024mapping, kousha2026much, matsui2025, geng_human-llm_2025}, and our 2024 to 2026 editions move in the same direction. But the same period, read against the full history, is the tail of a long trend rather than a break from a settled past. Two of these studies also show that adoption is uneven across fields and authors~\cite{lin2025divergentllmadoptionheterogeneous, kousha2024}, which is one reason a single short window can over- or understate the change depending on the venue it samples. A 44-year baseline lets us distinguish a genuine discontinuity in word choice from the acceleration of an already existing trend. Thus, LLMs may serve as a catalyst for ongoing prose change.

\subsection{Homogenization of Writing}
\textbf{RQ3} asked if writing became homogenized through the history of CHI. Mostly, it does not, and the three predictions fail in different ways. Vocabulary widened rather than narrowed. Length-matched MTLD rose 18.9\%, and Yule's K fell 12.7\%, and both hold on full body text as well as on the opening window. Sentence rhythm did not flatten. The coefficient of variation of sentence length is unchanged at delta 0.004, a shift in the third percentile of the history of CHI. Shared four-gram phrasing rose about a tenth above the preceding decade from 2021 to 2025, and the rise persists after removing artificial-intelligence vocabulary, so it is not just a shared topic. But it starts three editions before the first cohort that could have used these tools, and 2026 falls back into the historical band.

Our results contradict the claim that LLMs shrink linguistic diversity~\cite{sourati_shrinking_2026}. That claim, and related findings that these models reduce content and idea diversity~\cite{padmakumar2024, doshi2024, anderson2024}, mostly come from controlled generation or single-artifact studies, in which an LLM output is compared against a human baseline. When measured across the published CHI papers, we find the opposite sign in the diversity measures, so the homogenization prediction fares worst exactly where it should do best. The likely reason is that published papers are written by people who adopt these tools unevenly and edit their output~\cite{lin2025divergentllmadoptionheterogeneous, draxler2024}, so venue-level prose need not inherit the narrowing seen in raw model text. Our result does not contradict those studies on their own terms. It bounds how far their findings carry into the real literature. The register did not narrow, but densified, which is a different finding.

\subsection{The Rise and Retreat of AI-Typical Marker Words}
The AI-typical marker words behave unlike every other measure in our analysis. Words flagged by prior work as tells of generated text, \emph{delve}, \emph{intricate}, \emph{realm}, \emph{showcase}, and \emph{unveil}, among them, spiked into 2024 and then fell back. Eleven of the seventeen markers more than doubled against their 2021 to 2022 baseline by 2024, but only six still did by 2026, and seven of the eleven dropped below baseline entirely. Delve is the clearest case showing a spike in 2024, then a decline by 2026. This is the reverse of every register and diversity result, which do not turn around in 2026.

One possibility is that authors pulled back from using these words when it became clear that they are common AI-typical marker words~\cite{comasforgas2026}. We do not think this is what happened, because our data shows a large retreat from some words, but not from all of them. A plausible explanation is that the LLMs stopped producing specific AI-typical marker words and were less likely to be used in generated text. The same words that prior work identified as giveaways~\cite{comasforgas2025, comasforgas2026, kobak2025, matsui2025} became public knowledge, and the model versions widely used by 2026 no longer emit them at the earlier rate. The tell was removed while the assistance continued. This fits accounts of human and machine writing coevolving, in which each side adjusts to the other rather than one simply stamping its style onto the other~\cite{geng_human-llm_2025, lin2025divergentllmadoptionheterogeneous}. Our data support the conclusion that the 2024 spike is an artifact of a single model generation, not a stable feature of AI-assisted prose.

The consequence is methodological and runs counter to a growing practice. Marker-word lists are a moving target. A detector calibrated on the vocabulary of 2023 and 2024 output will under-count 2026 output even if real use is flat or rising, because the vocabulary it keys on has already been tuned away~\cite{liang2024mapping, kousha2026much}. This is why our main analysis does not rest on these markers. The stylistic signal we report, denser word choice and wider vocabulary, is measured on features no one was trying to hide, and unlike the markers, it does not reverse when the model behind the writing changes.

\subsection{Reading Ease Fell since CHI 2021}
In 2022, CHI crossed from the band the Flesch scale calls difficult into the one it calls very difficult, and by 2026, it resulted in 18.6\footnote{Scores below 30 are considered as ``Very Difficult'' text.}. This aligns with the one prior study that tracked readability across the ChatGPT transition, which found that scientific text became harder to read rather than easier~\cite{alsudais_exploring_2025}. The mechanism is visible in the function words. The sharpest single feature separating recent papers from older ones is the fall of \emph{of}, with \emph{a}, \emph{an}, \emph{is}, and \emph{are}, the signature of phrasal compression~\cite{BIBER20102}. Modification moves inside the noun phrase, so \emph{adaptation of the user interface} becomes \emph{user interface adaptation}, and a reader has to unpack a noun stack the author never had to build on purpose. CHI prose has become more personal and less penetrable at once. It is no longer anonymous. It is denser, and density is what readers pay for. However, this can be due to the removal of page limits in 2021. CHI removed fixed page limits in 2021 and moved to a contribution-defined length, and paper length has risen since. The word-level break and the steepest part of the reading-ease decline fall in that same window. We cannot separate the format change from any tool effect with these data, and we do not claim to, but a venue that stopped rewarding brevity is part of the context for prose that grew longer and denser.

\subsection{LLMs as a Catalyst for Writing}
Arguing against these tools would be both futile and unsupported by our data. The word-level break fell in 2021, and those papers were submitted in September 2020, more than two years before ChatGPT was public. The decline in reading ease is visible decade by decade since the 2000s. Whatever is compressing CHI prose was running long before a model could be blamed for it, which fits evidence that human and machine writing have been converging from both sides rather than one overwriting the other~\cite{geng_human-llm_2025}.

What the tools appear to have done is speed it up. The steepest single-year fall in reading ease in the corpus is 2025 to 2026, at 5.0 points, larger than the whole 2.9-point drift from 1982 to 2020. A system that produces fluent, dense, register-appropriate prose on demand lowers the cost of writing the kind of sentence the field was already drifting toward, which is the same lever the writing-assistant literature has shown can also steer content and opinion~\cite{jakesch2023, lee2022coauthor}. That is what a catalyst does. It sets the pace, not the direction. This reframes the practical question. The useful question is not whether to use these tools, which is settled in practice~\cite{salvagno2023, hutson2022}, but what to ask them for. Asked to raise the register, they push prose further along the trajectory we document. Asked to lower it, they are the fastest readability check the field has ever had.

\section{Conclusion}
We analyzed the full papers from 44 CHI proceedings to investigate whether LLMs are flattening academic prose into a single generic register. Our findings show that vocabulary widened rather than narrowed, sentence rhythm did not grow more uniform, and the one prediction that held, a modest rise in shared phrasing, began three editions before any CHI author could have used these tools. What we found instead is that CHI writing changed more between 2016 and 2026 than in any decade on record, and the direction is density. Papers grew fourfold over the full period, while reading ease fell from 37.7 in the 1990s to 18.6 in 2026, according to the Flesch reading ease. We found evidence that lexical trends began earlier at various paper-change events, for example, with the removal of page limits at CHI 2021. However, we state that LLMs accelerate the trends found in our measures. That leaves the community with a choice rather than a verdict. The same tools that make dense prose cheap to produce make legible prose cheap to produce, and which one the field gets depends on what it asks for and what it rewards. Scientific writing exists to move knowledge between people, and a literature that grows measurably harder to read each year is a field slowly raising the cost of its own understanding.

\bibliographystyle{ACM-Reference-Format}
\bibliography{references}
%TC:ignore
\appendix

\section{Preprocessing and Measure Definitions}
\label{app-preprocessing}

This appendix states the choices that determine every number in the paper. They are reported in full because tokenization and word-list membership change results quietly.

\subsection{Tokenization}

A token is matched by the regular expression \[\texttt{[A-Za-z][A-Za-z'\char39\char45]*}\] that is, a letter followed by any number of letters, apostrophes or hyphens. Three consequences follow. Punctuation is never a token, so the words \emph{word} and \emph{token} are interchangeable everywhere in this paper. Clitics stay attached, so \emph{Tom's} is one token rather than two. Hyphenated compounds stay whole, so \emph{human-computer} is one token, which matters for a corpus in which intra-word hyphenation rises by half over the period studied. Numerals and standalone symbols are not tokens and therefore do not enter any rate.

Text is lowercased before counting. No stemming or lemmatization is applied anywhere except in the AI-marker replication, where the published marker lists are defined over stems, and matching them requires stem matching. Stemming is avoided elsewhere because inflection is part of what we measure.

\subsection{Sentence segmentation}

A sentence boundary is terminal punctuation, one or more of \texttt{.!?}, optionally followed by closing quotation marks or brackets, then whitespace. A colon does not end a sentence. The following 32 abbreviations and initials are protected so that they do not create spurious boundaries.

\begin{quote}\small
al, approx, avg, cf, ch, dr, e.g, eq, et al, fig, figs, i.e, inc, jr, ltd, max, min, mr, mrs, ms, no, pp, prof, ref, refs, resp, sec, sr, st, std, vol, vs
\end{quote}

\subsection{Word lists}

Hedges, boosters, self-mention, and modals are counted by membership of the fixed lists below. Membership is the whole test. No attempt is made to judge what a marker is doing in its sentence, so these measures record the availability of a stance marker rather than its function, and a hedge inside a quotation counts the same as one in the authors' own voice.

\paragraph{Hedges (57 items).}
\begin{quote}\small
apparently, appear, appeared, appears, approximately, arguably, around, assume, assumed, assumes, believe, believes, could, estimate, estimated, fairly, generally, indicate, indicated, indicates, largely, likely, may, might, mostly, often, partly, perhaps, plausible, possible, possibly, potentially, presumably, probable, probably, quite, rather, relatively, roughly, seem, seemed, seems, somewhat, suggest, suggested, suggesting, suggestive, suggests, suppose, supposed, tend, tended, tends, typically, uncertain, unclear, usually
\end{quote}

\paragraph{Boosters (41 items).}
\begin{quote}\small
always, certainly, clearly, conclusively, considerably, critical, crucial, decisively, definitely, demonstrate, demonstrated, demonstrates, essential, establish, established, establishes, evidently, extremely, fundamental, highly, indeed, key, major, markedly, must, never, notably, obviously, plainly, prove, proved, proves, show, showed, shown, shows, significantly, strongly, substantially, undoubtedly, vital
\end{quote}

\paragraph{Self-mention (8 items).}
\begin{quote}\small
i, me, my, our, ours, ourselves, us, we
\end{quote}

\paragraph{Modals (10 items).}
\begin{quote}\small
can, could, may, might, must, ought, shall, should, will, would
\end{quote}

\paragraph{Function words (161 items).}
Lexical density is the share of tokens absent from this list, so it is the exact complement of the function-word share.
\begin{quote}\small
a, about, above, across, after, again, against, along, already, also, although, am, among, an, and, another, any, are, around, as, at, be, because, been, before, behind, being, below, beneath, beside, between, beyond, both, but, by, can, could, despite, did, do, does, doing, during, each, either, even, every, except, for, from, had, has, have, having, he, her, here, hers, herself, him, himself, his, how, i, if, in, inside, into, is, it, its, itself, just, least, less, may, me, might, mine, more, most, must, my, myself, neither, no, nor, not, of, on, only, onto, or, other, ought, our, ours, ourselves, outside, over, own, past, same, shall, she, should, since, so, some, still, such, than, that, the, their, theirs, them, themselves, then, there, these, they, this, those, though, through, to, too, toward, towards, under, underneath, unless, until, upon, us, very, was, we, were, what, when, where, whereas, whether, which, while, who, whom, whose, why, will, with, within, without, would, yet, you, your, yours, yourself
\end{quote}

The period classifier uses a slightly different closed-class list of 157 items. It drops the reflexive and possessive forms \emph{hers}, \emph{herself}, \emph{himself}, \emph{mine}, \emph{myself}, \emph{ours}, \emph{theirs}, \emph{yours} and \emph{yourself}, and adds the quantifiers \emph{all}, \emph{few}, \emph{many}, \emph{much} and \emph{several}. The two lists were built for different purposes, and we report both rather than implying one.

\subsection{Derived measures}

Nominalization is detected by suffix, using \texttt{\seqsplit{\textbackslash w\{4,\}(tion|sion|ment|ness|ity| ance|ence|ism|ship)s?\$}}, so a word of at least four characters ending in one of those suffixes counts, with an optional plural. Passive voice is a form of \emph{be} followed within two tokens by a word ending in \emph{ed} or a listed irregular participle. Both are approximations, and both are applied identically across all 44 years, so any error they carry is constant rather than year-varying. Rates are per 10,000 tokens for register measures and per 1,000 tokens for punctuation, computed against the body-text token count of the document in question.
%TC:endignore
\end{document}